\documentclass[a4paper,fleqn,usenatbib,useAMS]{mnras}

\usepackage{graphicx}	% Including figure files
\usepackage{amsmath}	% Advanced maths commands
\usepackage{amssymb}	% Extra maths symbols
\usepackage{multicol}        % Multi-column entries in tables
\usepackage{bm}		% Bold maths symbols, including upright Greek
\usepackage{pdflscape}	% Landscape pages
\usepackage{soul} %required by \st command
\usepackage{auto-pst-pdf}
\usepackage[T1]{fontenc}
\usepackage{ae,aecompl}

\usepackage{newtxtext,newtxmath}
\title[The galaxy cluster RXC J1504-0248]{A Gemini view of the galaxy cluster RXC J1504-0248: insights on the nature of the central gaseous filaments}

\author[A. C. Soja]{A. C.~Soja,$^{1}$\thanks{Contact e-mail: \href{mailto:ac.soja@gmail.com}{ac.soja@gmail.com}}
  L.~Sodr\'{e} Jr,$^1$ 
  R.~Monteiro-Oliveira,$^{2,1}$
  E. S.~Cypriano$^1$ and
  \newauthor 
  G. B.~Lima Neto$^1$
\\
  $^1$Universidade de S\~ao Paulo, Inst. de Astronomia, Geof\'isica e Ci\^encias Atmosf\'ericas, Depto. de Astronomia, R. do Mat\~ao 1226, 05508-090 S\~ao Paulo, Brazil\\
  $^2$Universidade Federal do Rio Grande do Sul,  Instituto de F\'isica, Departamento de Astronomia,  Campus do Vale, 91501-970 Porto Alegre, Brazil\\
  }

\date{Accepted 2018 March 07. Received 2017 December 20; in original form 2017 May 04}

\pubyear{2017}

\begin{document}
\label{firstpage}
\pagerange{\pageref{firstpage}--\pageref{lastpage}}
\maketitle

\begin{abstract}
We revisit the galaxy cluster RXC J1504-0248, a remarkable example of a structure with a strong cool core in a near redshift ($z = 0.216$). We performed a combined analysis using photometric and spectroscopic data obtained at Gemini South Telescope. We estimated the cluster mass through gravitational lensing, obtaining $M_{200} = 5.3\pm0.4 \times 10^{14}$ $h_{70}^{-1}$ M$_{\odot}$ within  $R_{200} = 1.56 \pm 0.04$ $h^{-1}_{70}$ Mpc, in agreement with a virial mass estimate. This cluster presents a prominent filamentary structure associated to its BCG, located mainly along its major axis and aligned with the X-ray emission. A combined study of three emission line diagnostic diagrams has shown that the filament emission falls in the so-called transition region of these diagrams. Consequently, several ionizing sources should be playing an meaningful role. We have argued that old stars, often invoked to explain LINER emission, should not be the major source of ionization. We have noticed that most of the filamentary emission has line ratios consistent with the shock excitation limits obtained from shock models. We also found that line fluxes are related to gas velocities (here estimated from line widths) by power-laws with slopes in the range expected from shock models. These models also show, however, that only $\sim$10\%  of H$\alpha$ luminosity can be explained by shocks. We conclude that shocks probably associated to the cooling of the intracluster gas in a filamentary structure may indeed be contributing to the filament nebular emission, but can not be the major source of ionizing photons.
\end{abstract}

\begin{keywords}
clusters: individual: RXC J1504-0248 -- clusters: ICM -- gravitational lensing: weak -- gravitational lensing: strong
\end{keywords}

%%%%%%%%%%%%%%%%%%%%%%%%%%%%%%%%%%%%%%%%%%%%%%%%%%

%%%%%%%%%%%%%%%%% BODY OF PAPER %%%%%%%%%%%%%%%%%%

\section{Introduction}
\label{sec_intro}

Currently, one of the greatest challenges in galaxy cluster physics  is  understanding the mechanisms that regulate the hot intracluster gas (ICM) thermal properties. It is expected from simple models \citep[see][for a review]{Sarazin88} that for massive clusters the gas could cool at rates of thousands of solar masses per year. Such flows, however, have never been observed at that intensity \citep[e.g.,][]{Fabian03} and, then, one or more mechanisms are needed to explain the heating of the gas.  Many of them have been proposed, like active galactic nucleus (AGN) feedback \citep[e.g.,][]{Blanton01, McNamara12, Gaspari12}, gas dynamics \citep[e.g.][]{Semler12, McDonald12}, events like cluster mergers \citep[e.g.][]{McGlynn84, ZuHone10} or gas heating during the early stages of cluster formation \citep[e.g.][]{Kaiser91, McCarthy08}. Despite the AGN feedback being considered almost a paradigm as the main heating source (as well as for quenching or triggering star formation), recent studies \citep[e.g.][]{Fabian17, Hillel17} have shown that other processes could play an important role to keep cooling flows at their observed rates.

One way to study in more detail the mechanisms which control the ICM heating is to investigate clusters where cooling flows are expected, but not observed, at least at the rates expected from simple models. Of particular interest is to examine the dynamical status of galaxy clusters to verify the role of equilibrium in keeping the cool core phenomenon. Indeed,  as argued by \citet{Edu04}, a comparison of masses obtained by techniques which assume some sort of dynamical equilibrium (like those determined from the cluster X-ray emission or by applying the virial theorem to galaxy motions) with those which are free from this assumption (e.g. gravitational lensing analysis) is a powerful tool for probing the dynamical state of a cluster.

Another approach is to analyse the nature of possible heating sources by the expected effects on the line emission of the gas in the central cluster galaxy. To this purpose, emission line diagnostic diagrams are often adopted. The most widely used diagnostic diagram is the [OIII]5007/H${\beta}$ versus [NII]6584/H$\alpha$, the BPT diagram \citep{Baldwin81}, which classifies the excitation source into either star-formation or AGN. However, several works have shown that different phenomena can occupy similar areas in the BPT diagram, mainly in the transition region of this diagram. Examples include old stars  \citep[e.g.][]{Grazyna08} and shocks \citep[e.g.][]{Dopita94,Alatalo16,Herpich16}, so a deeper analysis considering different models is necessary to understand what kind of source dominates nebular excitation.

The galaxy cluster RXC J1504-0248 (hereafter RX1504), at $z=0.216$, is very massive  and is considered the most prominent  cool-core cluster in the local Universe \citep{Bohringer05}, due to its very high formal mass deposition rate estimated assuming the classical cooling flow model (see below). Fig.~\ref{cortes} depicts its  central region in optical wavelengths, showing its brightest cluster galaxy (BCG) and some gravitational arcs. The BCG is an interesting object by itself, presenting several filaments with emission attributed to ongoing star-formation \citep{Ogrean10}. There are interesting investigations about this cluster in several wavelength ranges, but when compared to other remarkable cool core clusters, such as Perseus \citep[e.g.,][]{Conselice01,Churazov04,Hatch05,Salome06,Canning14}, it is still understudied.

In one of the first investigations dedicated to RX1504, \cite{Bohringer05}, using X-ray data from the Chandra satellite, determined that this cluster  is very hot and luminous: $ kT = 10.1$ keV, $L_X = 4.3 \times 10^{45}$ erg s$^{-1}$ (for $h_{100}=0.7$). They noticed that it is the most luminous cluster in the REFLEX survey \citep{Bohringer01}. 
By fitting a beta-model to the X-ray emission, they obtained an estimate of the mass deposition rate in the classical cooling flow model of $ \sim 1500 - 1900$ M$_{\odot}$ yr$^{-1}$. They also determined the  cluster total mass, obtaining $ 1.7 \times 10^{15} h _{70} ^{-1}$ M$_{\odot} $ within $ 3 h_{70}^{-1} $ Mpc, i.e., it is very massive.  With the same data and through a hidrodynamical analysis, \cite{Zhang12} determined the cluster total mass inside a $ 1.18~  h_{70}^{-1} $ Mpc radius as $ 5.4 \times 10^{14} ~ h _{70} ^{-1}$ M$_{\odot}$. 

Investigating the optical region, \cite{Ogrean10} found a cooling flow mass deposition rate ranging between 136 and 239 $h^{-1}_{71}$ M$_{\odot}$ yr$^{-1}$, corresponding to, approximately, less than 10 per cent of the mass deposition rate expected from the total X-ray luminosity emitted from the cluster core. Using a BPT diagram, they concluded that stellar UV is a plausible ionizing source, but cannot act alone. \cite{Mittal15} revisited the star formation rate (SFR) in RX1504 through a Bayesian analysis using stellar synthesis and found a lower value,  $67 \substack{+49 \\ -27} ~  h^{-2} _{71}$ M$_{\odot}$ yr$^{-1}$, which reinforces the cooling flow problem. 

Additionally, \cite{Giacintucci11} discovered in this cluster a diffuse radio source, classified as a mini halo. This source was detected using the Giant Metrewave Radio Telescope at 327MHz, and confirmed with data from the VLA 1.46GHz. This kind of structure is largely found in strong cool core clusters \citep[e.g. Perseus,][]{Fabian06}, although it is still a rare phenomenon, since there are only $\sim$ 20 radio mini halos known  \citep[][]{Bravi15}.

In this paper we revisit RX1504 through an optical analysis using Gemini South data. We have two main objectives. The first is to
investigate whether it is in dynamical equilibrium by comparing a virial mass estimate, which is sensitive to departures of dynamical 
equilibrium, with the results of a gravitational lensing analysis, which is not dependent of the cluster dynamics \citep[][]{Edu04}.
Our second main goal is to discuss the  main excitation sources of the gas emission in the filaments around the BCG through 
the study of emission line ratios. 

The structure of the paper is as follows. In section 2 we describe the Gemini observations and data reduction. In section 3 we present our estimates for the cluster mass through velocity dispersion and weak lensing analysis, and compare our results with previous estimates. In section 4 we discuss the nature of the gas emission coming from the cluster BCG. The main results of this paper are discussed in section 5 and summarized in section 6. Throughout the paper we assume a cosmology with $\Omega _m = 0.3$, $\Omega _{\Lambda} = 0.7$ and $H_0 = 70 ~h_{70}$ km s$^{-1}$ Mpc$^{-1}$. With these  cosmological parameters, the mean cluster redshift $\bar{z}=0.2164$ corresponds to a luminosity distance  equal to $d_L  = 1063.5$ Mpc and a scale of $3.5$ kpc/arcsec.

\section{Observations and Data Reduction}
\label{sec_data}

In this section we present the optical imaging and multi-slit spectroscopic observations of the galaxy cluster RXC J1504-0248. These observations were done with the GMOS instrument, mounted in the Gemini South Telescope. The observational details and the main data reduction steps are described in the sub-sections below. 

\subsection{Optical Imaging}
\label{ssec_imaging}

The photometric observations were done in May 25, 2009, under photometric conditions and in queue mode (GS-2009A-Q5; PI: L. Sodre). The GMOS field of view has $5.5 \times 5.5$ sq. arcmin and the detector pixel size is $0.1460$ arcsec. The imaging comprised three exposures of 200s each, in three filters of the Sloan Digital Sky Survey (SDSS) system:  $g^\prime$, $r^\prime$ and $i^\prime$  \citep{Fukugita96}. All basic steps (bias, dark, flat field and sky subtraction) of the image reduction were done with the GMOS package running under the {\tt IRAF} environment. 

\begin{figure}
\begin{center}
\includegraphics[width=0.45\textwidth,angle=0]{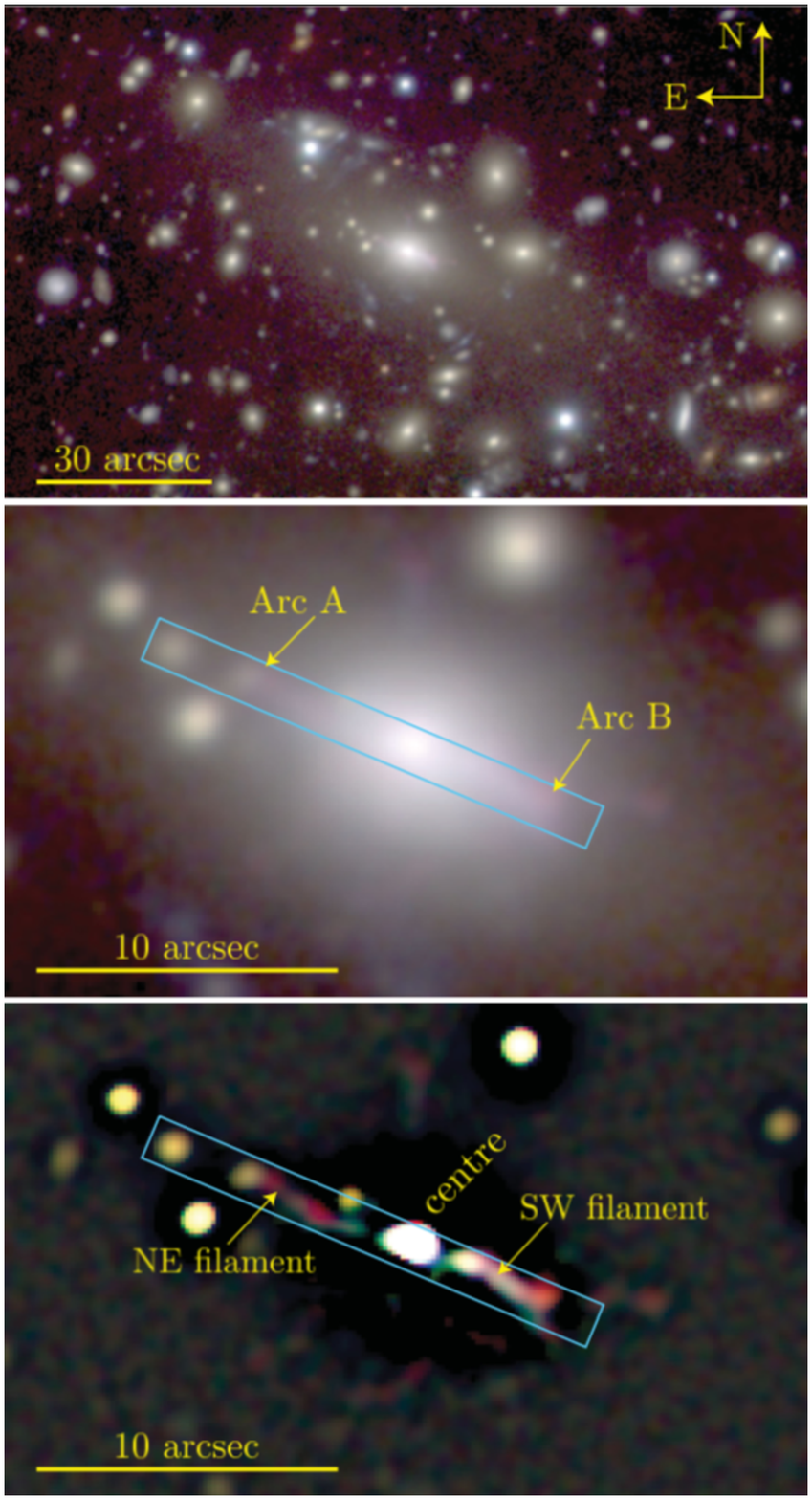}
\caption[]{Top: Composed $g^\prime$, $r^\prime$ and $i^\prime$ Gemini South images of the central region of the galaxy cluster RXC J1504-0248 (RXCJ1504, $\bar{z}=0.216$). Middle: Zoomed image from the BCG region showing the filamentary structure around it. A gravitational arc (named ``Arc A'') was discovered by our imaging analysis whereas its counterpart (``Arc B'') was identified through the BCG spectrum.  The position of the slit used for the spectroscopic analysis is also shown. Bottom: Unsharp masking image of the BCG region, highlighting the main filamentary structure along the NE--SW direction. This corresponds to the main UV emission \citep[as we can see in ][]{Tremblay15}. We named the three regions along the slit where we will carry further comparisons in our spectroscopic analysis (NE, centre and SW).}
\label{cortes}
\end{center}
\end{figure}

Fig. \ref{cortes} (top) depicts a composed image of the central region of the cluster. Fig. \ref{cortes} (middle) is a close-up of the central galaxy showing two gravitational arcs discussed in Sect. \ref{sec.grav.arc}. Fig. \ref{cortes} (bottom) is an unsharp masking image revealing the gaseous filament extending from NE to SW along the major axis of the central galaxy (see Sect. \ref{ssec_morphology}).

To create a catalogue of the objects in the field, we have used the {\tt SExtractor} code \citep[hereafter SE,][]{Bertin96} in dual mode, with the $r^\prime$-band set as the detection frame\footnote{We have adopted in our analysis SE {\sc MAG\_AUTO} magnitudes.}. The photometric calibration in the standard SDSS system was made through comparison between our data with SDSS objects in the same field. We found 172 galaxies in the SDSS database with coordinates in the range $226.00210 ^\circ < {\rm RA} <  226.03321 ^\circ$ and $-2.8514 < {\rm Dec} <  -2.8050$ (JD2000). To minimize photometric errors, we selected for calibration purposes only objects with SE instrumental magnitude error less than 0.1. The photometric calibration obtained with this procedure was then applied to the other galaxies in the field. The expected error in the zero points are $0.010$, $0.011$ and $0.007$ for $g^\prime$, $r^\prime$ and $i^\prime$, respectively. 

\begin{figure}
\begin{center}
\includegraphics[width=0.45\textwidth]{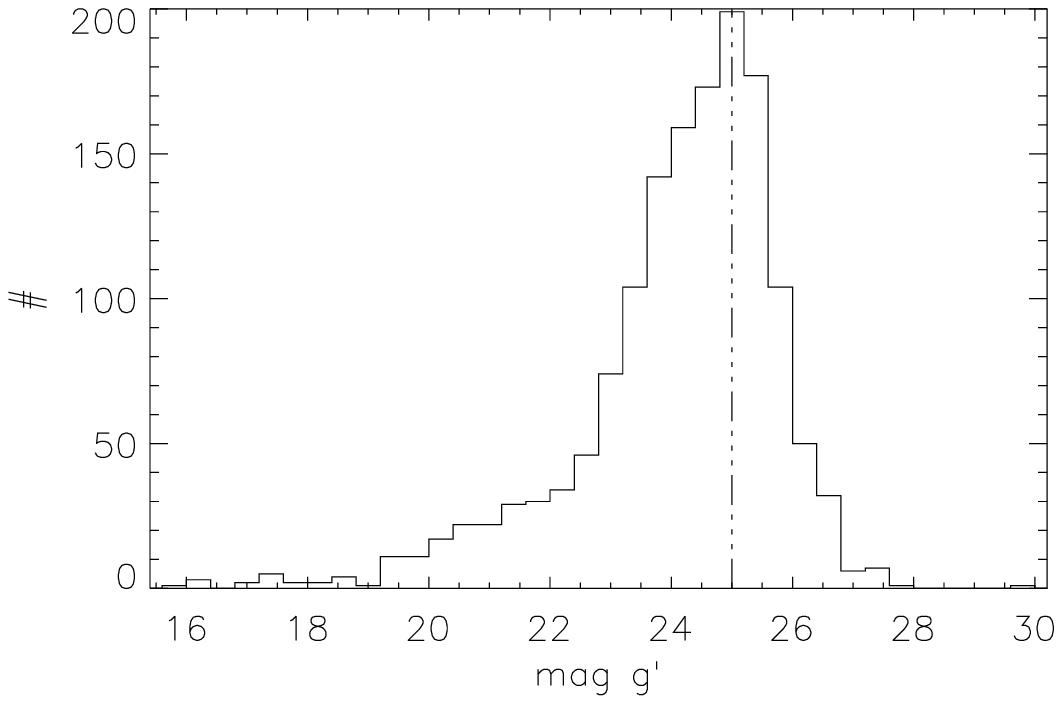}\quad
\includegraphics[width=0.45\textwidth]{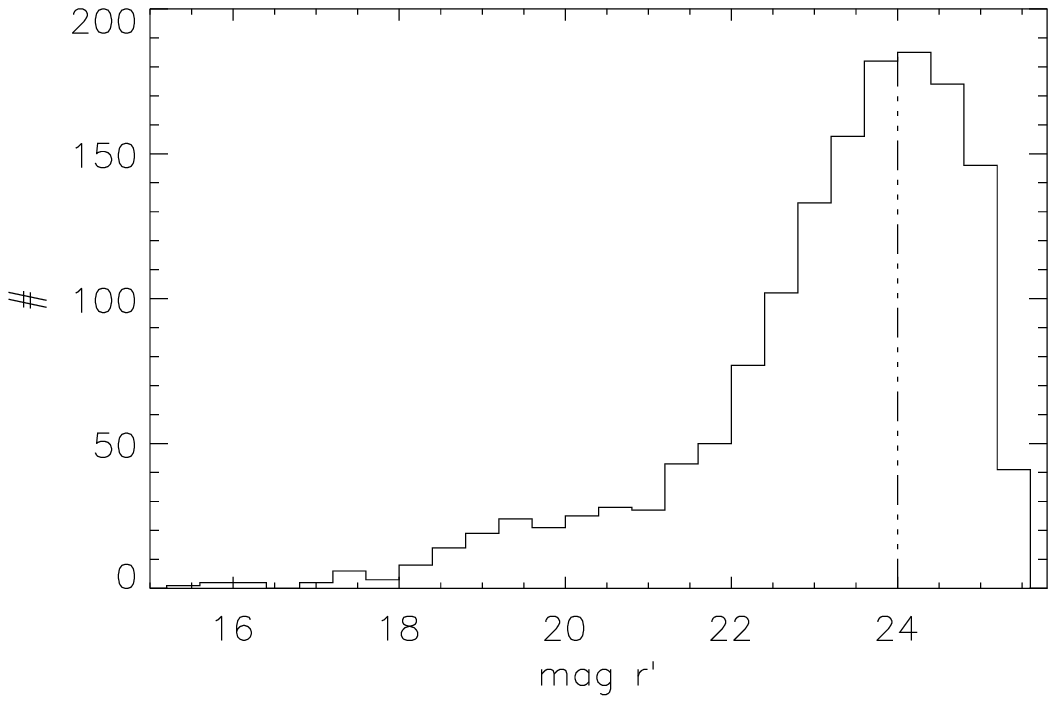}\quad
\includegraphics[width=0.45\textwidth]{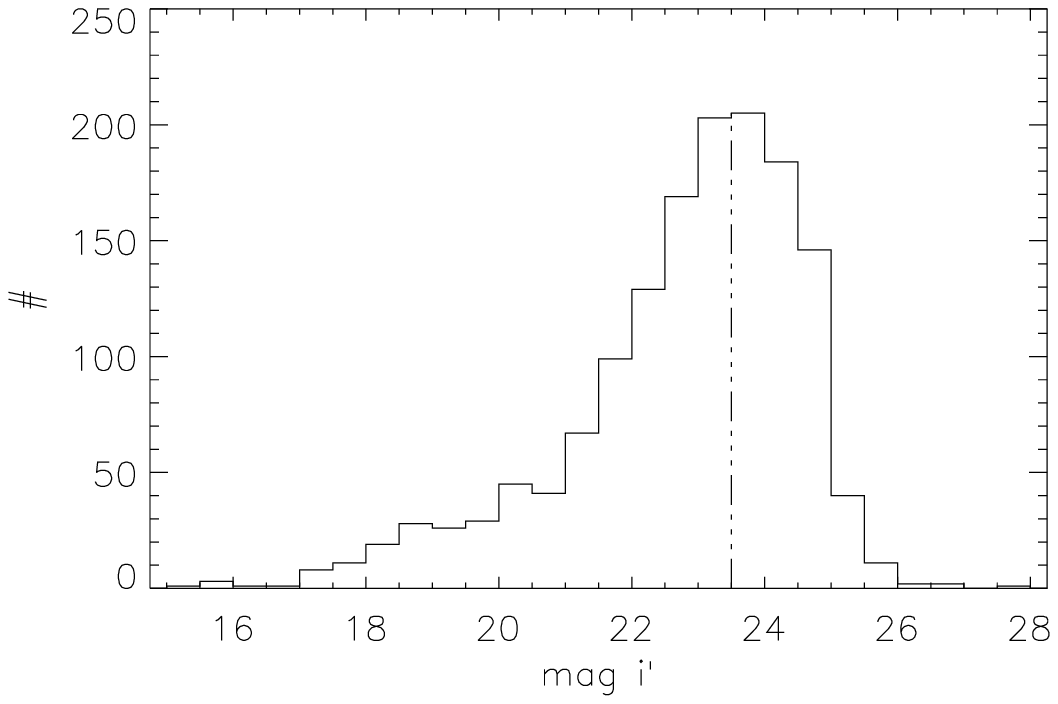}\quad
\caption[]{Magnitude distribution of galaxies in the field of RXC J1504-0248 for the $g^\prime$, $r^\prime$ and $i^\prime$ bands. The black vertical line shows the assumed limiting magnitude in each band.}
\label{compl}
\end{center}
\end{figure}

The magnitude limits for the photometric analysis were determined from the distribution of the number of objects as a function of the magnitude. We expect that the number of detected objects grows as the magnitude increases up to a certain value, before start decreasing (due to incompleteness), and adopt as the limit for this analysis the magnitude in each band where the counts are maximum: $24.50 \pm 0.02$, $24.00 \pm 0.04$ and $23.50 \pm 0.04$ for $g^\prime$, $r^\prime$ and $i^\prime$, respectively (Fig.~\ref{compl}). These results are consistent with values obtained in other works with Gemini South data using similar observational setups \citep[e.g.,][]{Carrasco07}. 

%<--Figure #1

%<--Figure #3

\begin{figure}
\begin{center}
\includegraphics[width=0.6\columnwidth,angle=-90]{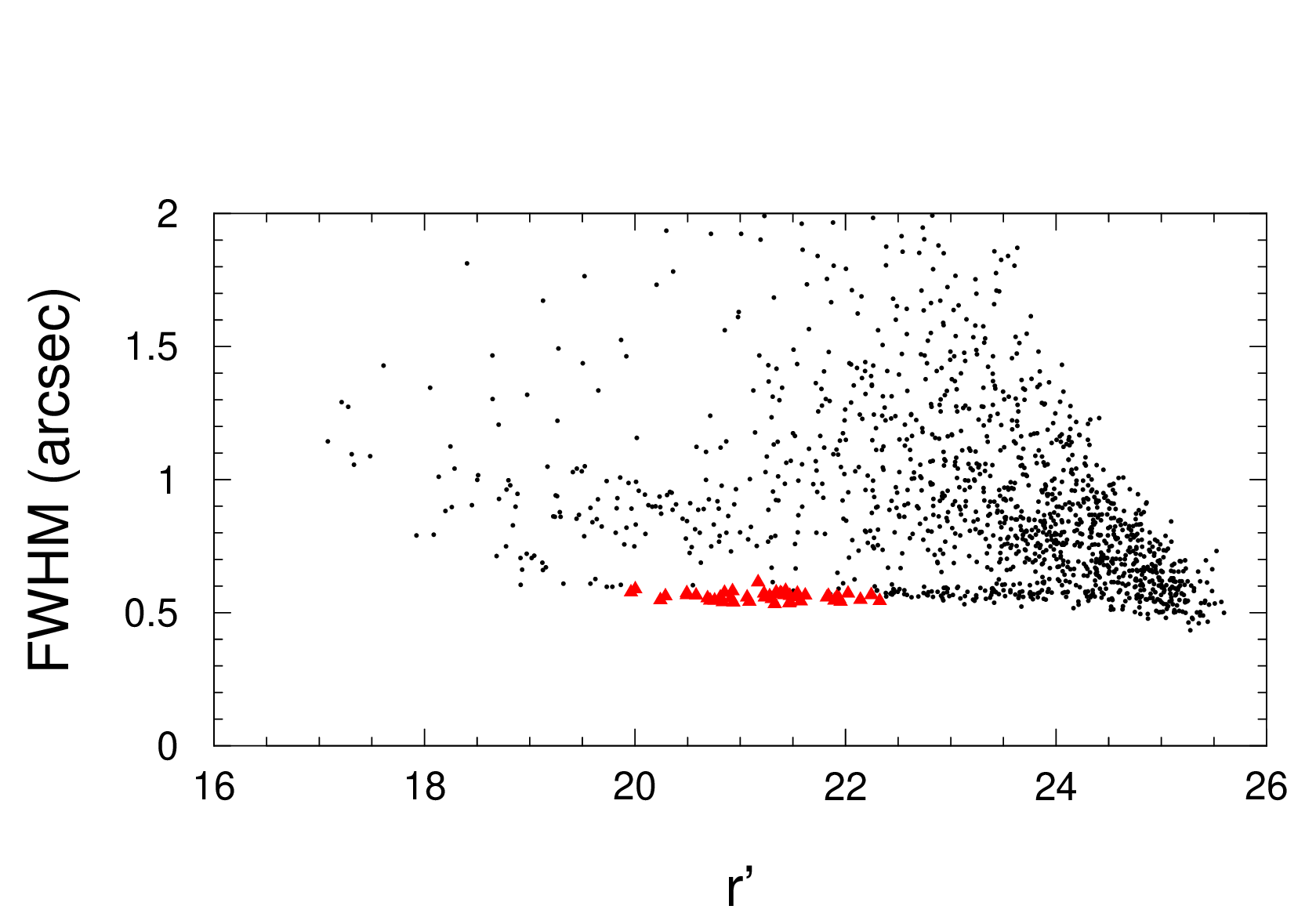}
\includegraphics[width=0.6\columnwidth,angle=-90]{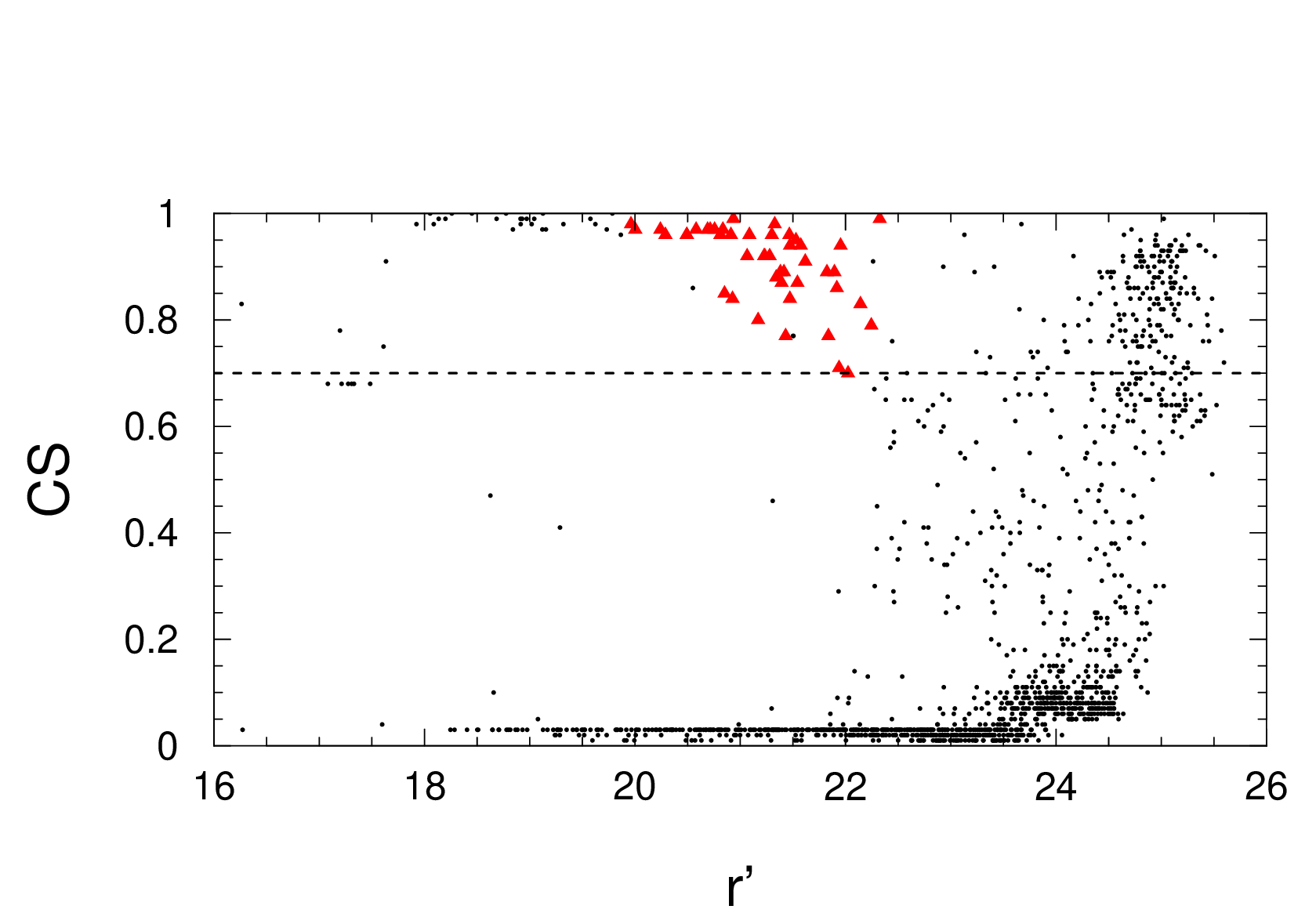}
\caption[]{{\it Top}: FWHM versus $r^\prime$ diagram of the photometric catalogue. In this plot, we clearly see a horizontal branch where the stars are preferentially located. The unsaturated stars were selected as the objects lying in the interval $20\leq r^\prime \leq 22.5$ (red triangles) and have a mean FWHM of 0.65 arcsec in the $r^\prime$ band. Complementarily, we have selected galaxy candidates as the objects having their FWHM larger than 5 pixels (0.73 arcsec) and r’=24 brighter than the limit magnitude in this band ($r'=24$). {\it Bottom}: Besides having their FWHW larger than the stars a second criterion was used to select the galaxies: their {\sc CLASS\_STAR} parameter. The red triangles correspond to the previously selected stars which have $CS=0.7$ as the minimum. This value was then adopted as the cutoff between the stars and galaxies (dashed black line). }
\label{estgal}
\end{center} 
\end{figure}

The weak-lensing analysis presented in Sect. \ref{ssec_lensing} requires the knowledge of the point spread function (PSF) across the field. The selection of unsaturated stars for this task was done based on the full width at half maximum (FWHM) values of the object images provided by SE in a FWHM versus magnitude diagram (Fig.~\ref{estgal}).  Non-saturated stellar objects are expected to form a horizontal sequence in this diagram, which is clearly seen in Fig.~\ref{estgal}. This figure shows that, to avoid saturation, we should select stars fainter than $r^\prime \simeq 20$. To assure a good S/N for the PSF determination, we have considered only objects brighter than $r^\prime = 22.5$. The stars selected in this way are shown as red triangles in the top panel of Fig.~\ref{estgal}. To check this selection we have examined the position of these stars in a diagram showing the SE stellarity parameter CS\footnote{Values of $CS$ close to one are expected for point sources, whereas values close to zero means that the object is definitely extended.} as a function of the magnitude. The bottom panel of Fig.~\ref{estgal} shows that all stars selected for the PSF analysis have CS larger than 0.7. Through a similar analysis in each band we have estimated the mean value of the FWHM as  0.78, 0.65 and 0.81 arcsec for the FWHM in the $g^\prime$, $r^\prime$ and $i^\prime$ images, respectively.

To identify the galaxies in the field we also made use of  Fig.~\ref{estgal}. Being conservative, we selected as galaxies for further analysis those objects with FWMH greater than 5 pixels (0.73 arcsec), with CS $<$ 0.7 and with $r^\prime < 24$, the limit magnitude in this band. Using this approach, we have identified 82 stars and 1216 galaxies in our field.

%<--Figure #3

\subsection{Optical Spectroscopy}
\label{ssec_spectra}
 
 The spectroscopic observations were performed in April 04, 2010, and July 08, 2010, under photometric conditions and in queue mode (GS-2010A-Q26; PI: L. Sodr\'e). The observations were done with two masks, using 1 arcsec slits and with the R400 grating. For each mask we obtained three exposures of 870 sec each. This set up gives spectra covering a range from $\sim$4000 \AA ~ to $\sim$8000 \AA ~  with resolution of $\sim 8$ \AA. In total, 67 slit spectra were obtained with these two masks.

We used the standard tasks in Gemini {\tt IRAF} package to reduce the data. The wavelength calibration was done through comparison with exposures of  $Cu-Ar$ lamps (we also use the lamps to determine the instrumental line widths). For the flux calibration we created a sensitivity function using the {\it standard } task from the {\it onedspec} {\tt IRAF} package applied to three observations of the standard star LTT 1788 \citep{Hamuy92}, which were also reduced using the same Gemini packages.

Radial velocities ($cz$) were determined using the {\tt IRAF} package RVSAO \citep{Kurtz98}, which implements the cross-correlation technique \citep{Tonry79} against a library of galaxy templates\footnote{\texttt{http://classic.sdss.org/dr7/algorithms/spectemplates/}}.  We obtained 42 useful galaxy spectra (with S/N $>$ 14, measured in the continuum region of $6300 - 6500$~\AA, or Tonry \& Davis $R$ statistics larger than $R = 2.5$). They include the BCG, cluster and field galaxies. For galaxies with prominent emission lines, radial velocities were measured using them when at least three emission lines were identified, in general the H$\alpha$ + [NII] or the H$\beta$ + [OIII] line sets. All the results were individually checked by eye.

The set of reliable redshifts is presented in Table \ref{tabelaxcsao}. We also included in this table additional 13 redshifts obtained using NED\footnote{The NASA/IPAC Extragalactic Database (NED) is operated by the Jet Propulsion Laboratory, California Institute of Technology, under contract with the National Aeronautics and Space Administration.} because they can be considered cluster members and will be useful for the analysis presented in the next section.

%<--Table #1
\begin{table*}
\begin{center}
\caption[]{Spectral data for galaxies in the field of the galaxy cluster RXC J1504-0248. The redshifts of objects with * were determined through emission lines ($H\alpha, H\beta$, and the $N[II], S[II]$ doublets). The $\dag$ highlights the BCG.}
\begin{tabular}{lcccccc}
\hline \hline 

{Observations} & {R.A.} & {DEC} & {S/N} & {R$_{\rm quality}$} & {z } & {Reference}\\

\hline 
HeCS		  & 225.96458	 &  -2.84722	 & -- & -- &	   0.21785   $\pm$   0.0000100	&(4)\\
HeCS		  & 225.98625	 &  -2.79250	 & -- & -- &	  0.2087940	$\pm$   0.00001	&(4)\\
GS 		  & 225.98820  &    -2.83871  &  27.98  &  2.80  &     0.21679 $\pm$       0.00101  & (1)      \\ %4 
HeCS		  & 225.98208	 &  -2.85694	 & -- & -- &	       0.21600	$\pm$   0.00001	&(4)\\
GS 		  & 225.99030  &    -2.83639  &  15.68  &  2.58   &    0.21233 $\pm$       0.00092  & (1)     \\ %32
GS 		  & 225.99046  &    -2.79717  &  58.59  &  4.99   &    0.21716 $\pm$       0.00014  & (1)      \\ %3
GS* 		  & 225.99297  &    -2.80465  &  14.10  &  3.04   &    0.12709 $\pm$       0.00013  & (1)      \\ %122
GS		  & 225.99420  &    -2.80742 &   20.62  &  2.57   &    0.13943 $\pm$       0.00039  & (1)     \\ %229
GS* 		  & 225.99568  &    -2.79738 &   17.95  &  3.32   &    0.12674 $\pm$       0.00016  & (1)     \\ %88
GS 		  & 225.99728  &    -2.83116 &   27.85  &  2.91   &    0.21789 $\pm$       0.00087  & (1)     \\ %65
GS 		  & 225.99829  &    -2.76596  &  20.64 &   4.22   &    0.21979 $\pm$       0.00016  & (1)     \\ %237
GS 		  & 225.99985   &   -2.77387 &   23.34  &  3.84   &    0.21710 $\pm$       0.00013  & (1)     \\ %426
GS 		  & 226.00020  &    -2.79121  &  30.51 &   2.59   &    0.11392 $\pm$       0.00035  & (1)     \\ %548
GS 		  & 226.00412   &   -2.75708  &  34.73  &  3.98   &    0.02422 $\pm$       0.00009  & (1)     \\ %650
GS 		  & 226.00528   &   -2.82123  &  58.59  &  3.76   &    0.21987 $\pm$       0.00080  & (1)     \\ %533
GS 		  & 226.00804   &   -2.81274  &  28.13  &  3.40    &   0.21695 $\pm$       0.00047  & (1)     \\ %747
GS		  & 226.01024   &   -2.75725  &  35.05  &  5.52   &    0.04128 $\pm$       0.00015  & (1)     \\ %920
GS 		  & 226.01169   &   -2.78995  &  34.56  &  2.87    &   0.21901 $\pm$       0.00059  & (1)     \\ %768
GS		  & 226.01239   &   -2.78909  &  48.89  &  4.21    &   0.21916 $\pm$       0.00014  & (1)     \\ %828
GS 		  & 226.01428   &   -2.76484  &  37.44  &  7.18    &   0.03594 $\pm$       0.00012  & (1)     \\ %1083
GS 		  & 226.01604   &   -2.84463  &  27.39 &   2.64    &   0.21900 $\pm$       0.00085  & (1)     \\ %958
GS 		  & 226.01926   &   -2.77116  &  27.15  &  3.02    &   0.21663 $\pm$       0.00022  & (1)     \\ %1161
GS 		  & 226.02046   &   -2.79257  &  37.67  &  4.86    &   0.21495 $\pm$       0.00016  & (1)     \\ %1204
GS 		  & 226.02155   &   -2.77161  &  48.89  &  4.28   &    0.02207 $\pm$       0.00014  & (1)    \\ %1321
HeCS		  & 226.02208	 &  -2.84944	  & -- & -- &	  0.22055	$\pm$   0.00001	&(4)\\
GS 		  & 226.02478   &   -2.75759  &  64.40 &   4.88    &   0.05517 $\pm$       0.00022  & (1)     \\ %1471
SDSS	          & 226.02583	 &  -2.80472	 & -- & -- &	  0.22100	$\pm$   0.00001	&(4)\\
GS 		  & 226.02704   &   -2.80097  &  49.80  &  5.24    &   0.20849 $\pm$       0.00014  & (1)     \\ %1315
HeCS		  & 226.02708	 &  -2.78250	 & -- & -- &	  0.21818	$\pm$   0.00001	& (4)\\
SDSS	          & 226.02916	 &  -2.81055	  & -- & -- &     0.21600	$\pm$   0.00001	&(4)\\
SDSS		  & 226.03041	 &  -2.80277	 & -- & -- &	  0.21510	$\pm$   0.00001 &	(2) \\
LCRS		  & 226.03125	 &  -2.80472	 & -- & -- &	  0.21707	$\pm$   0.00001	&(3)\\
 GS$^\dag$        & 226.03140    &  -2.80454  &  88.30  &  7.90    &   0.21571 $\pm$       0.00014  & (1)     \\ %9001
SDSS	          & 226.03333	&   -2.80444	 & -- & -- &	  0.21760	$\pm$   0.00001&	(2) \\
GS 		  & 226.03625 &     -2.76002 &   22.00 &   4.42  &   0.02283 $\pm$       0.00009  & (1)     \\ %1976
HeCS		  & 226.03708	 &  -2.85722	  & -- & -- &	  0.21784	$\pm$   0.00001	&(4)\\
GS		  &  226.04248  &    -2.78164  &  47.90 &   3.03   &   0.21491 $\pm$       0.00068  & (1)     \\ %3049
GS		  &  226.04405  &    -2.79589  &  50.96 &   4.79   &   0.21617 $\pm$       0.00018  & (1)     \\ %2042
GS* 		  &  226.04514  &    -2.82149  &   4.06 &   3.54   &   0.32100 $\pm$       0.00016  & (1)     \\ %2781
GS 		  &  226.04655  &    -2.77514  &  24.81 &   3.35   &   0.19442 $\pm$       0.00020  & (1)     \\ %2861
GS		  &  226.04800  &    -2.78937  &  19.05 &   2.84    &  0.21876 $\pm$       0.00063  & (1)     \\ %2841
 GS 		  &  226.04837  &    -2.81817  &  45.43 &  15.67   &   0.07530 $\pm$       0.00005  & (1)     \\ %2708
 GS 		  &  226.04967  &    -2.75773  &  20.38 &   8.25   &   0.03090 $\pm$       0.00095  & (1)     \\ %2780
 GS 		  &  226.05333  &    -2.80072   & 61.74 &   6.43   &   0.21402 $\pm$       0.00012  & (1)     \\ %2617
 GS 		  &  226.05617  &    -2.76181  &  37.88 &   4.93    &  0.03946 $\pm$       0.00009  & (1)     \\ %2527
 GS		  &  226.05844  &    -2.84156  &  79.72 &   2.72    &  0.12655 $\pm$       0.00023  & (1)     \\ %2320
 GS		  &  226.06241  &    -2.80750  &  43.18  &  6.06   &   0.21524 $\pm$       0.00012  & (1)     \\ %2236
 GS 		  &  226.06459  &    -2.75706  &  88.65  &  4.48   &   0.04914 $\pm$       0.00017  & (1)      \\ %2158
 GS	          &  226.06654  &    -2.79334  &  43.13  &  5.80   &   0.22111 $\pm$       0.00015  & (1)     \\ %3632
 GS		  &  226.06898  &    -2.78012  &  18.50 &   3.22   &   0.52999 $\pm$       0.00026  & (1)     \\ %3601
 GS 		  &  226.07195  &    -2.78775  &  14.00  &  4.13   &   0.21370 $\pm$       0.00014  & (1)      \\ %3520
 GS		  &  226.07196  &    -2.83000  &  24.40  &  2.53   &   0.73100 $\pm$       0.00021  & (1)     \\ %3310
 GS		  &  226.07439  &    -2.81395  &  56.45 &   2.53   &   0.21540  $\pm$      0.00005  & (1)     \\ %3369
 HeCS		  &  226.09250	 & -2.866388	 & -- & -- &	  0.21127	$\pm$   0.00001	&(4)\\
 HeCS		  &  226.09750	 & -2.827777	  & -- & -- &	 0.20889	$\pm$   0.00001	&(4)\\

%Arc \#1	         &   226.03441  &    -2.78491  &  19.7   &  6.95   &   1.21732 $\pm$       0.00114  & (1)   \\ %
Arc~A          &   226.03209  &    -2.80044  &  15.29 &   3.87   &  0.66334 $\pm$       0.00044  & (1)     \\ %
Arc~B          &   226.03209  &    -2.80044  &  15.29 &   3.87   &  0.66337 $\pm$       0.00044  & (1)     \\ %	

\hline 

\end{tabular}
\label{tabelaxcsao}

{References: (1) GS - Gemini South, this work, (2) SDSS - Sloan Digital Sky Survey,  \cite{Eisenstein11},(3) LCRS - Las Campanas Redshift Survey, \cite{Shectman96} and (4) HeCs - Hectospec Cluster Survey , \cite{Rines13}.}
\end{center}
\end{table*}

\begin{figure}
\begin{center}
\includegraphics[width=0.8\columnwidth,angle=270]{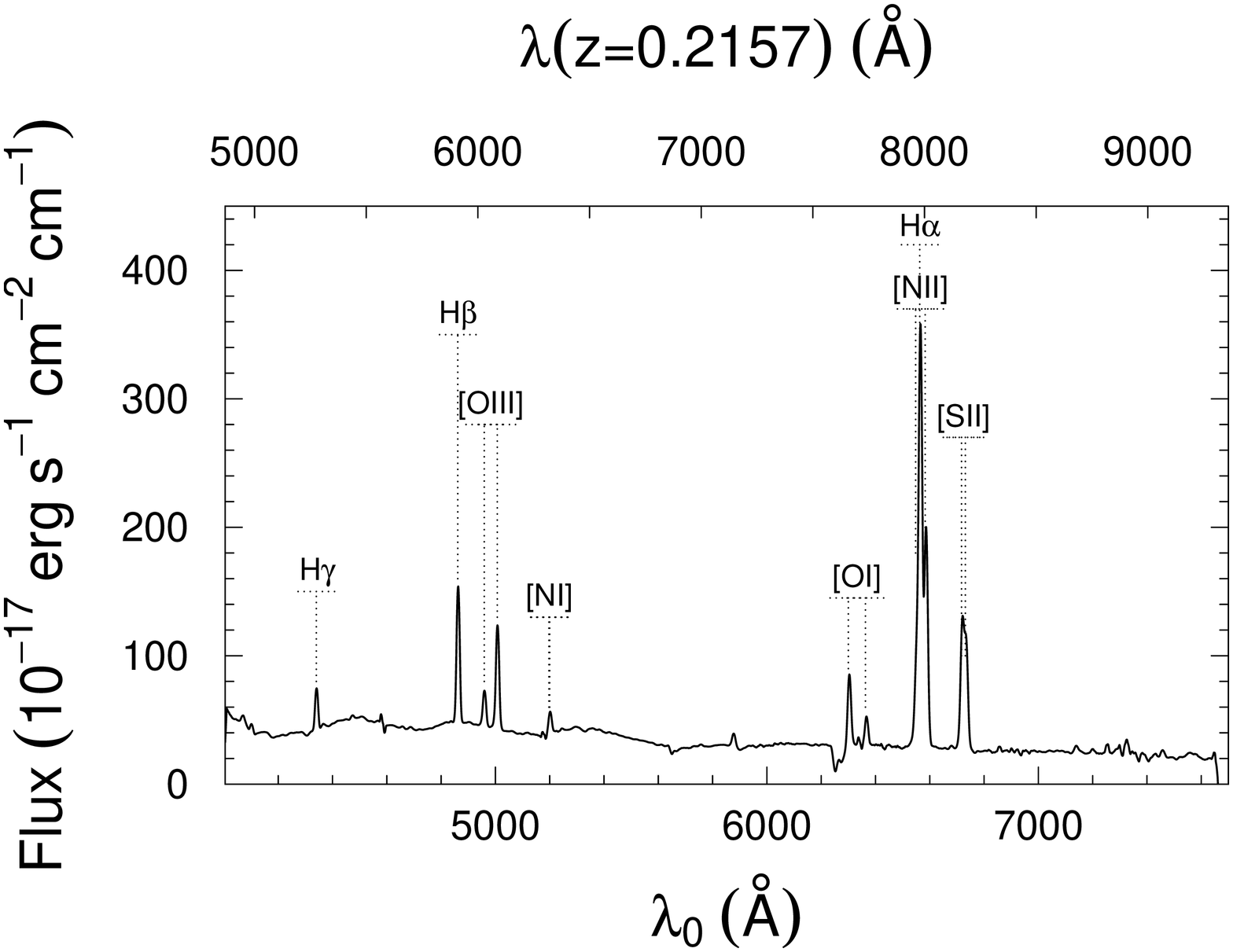}\quad
\includegraphics[width=\columnwidth]{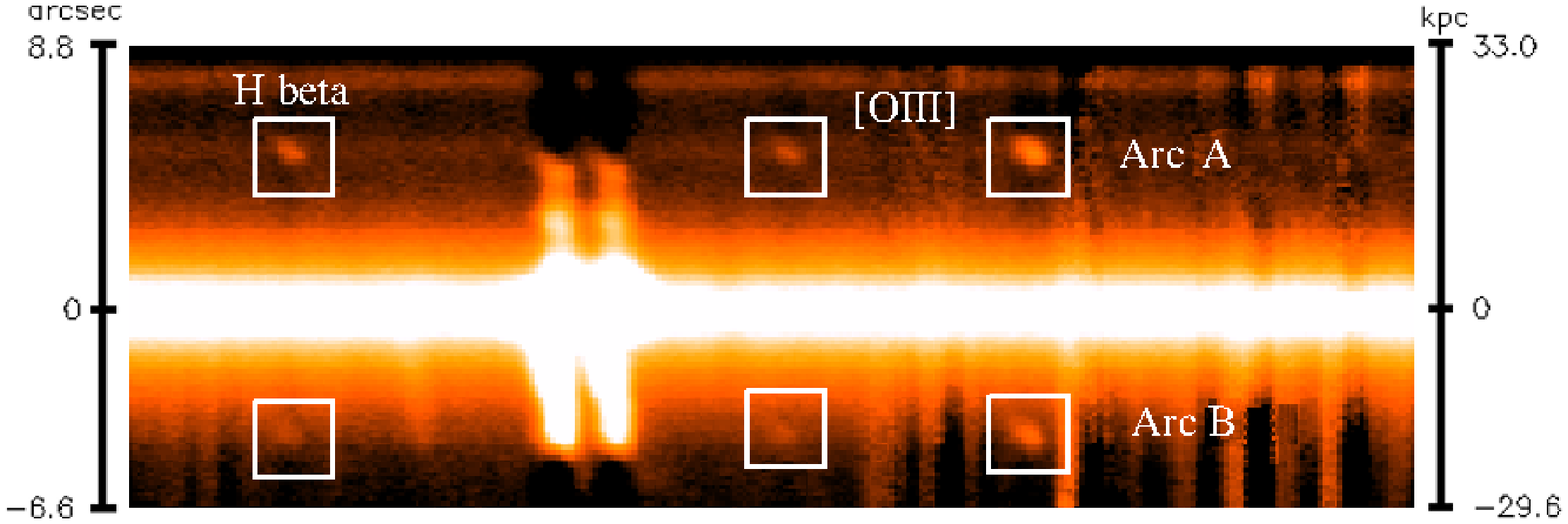}
\caption[]{Top: Integrated slit spectrum of the BCG. The dashed lines refer to significant spectral features. Bottom: the red part of the BCG spectrum inside the slit (we choose this part to highlight the emission lines). At the left side we show the intervals corresponding to the spatial segmentation of this spectrum in arcsec while in the right we show the same intervals in kpc. The rectangles highlight emission lines in both arc spectra. }
\label{espectros}
\end{center}
\end{figure}

One of the main motivations of this paper is to analyse the nature of the emission lines of the BCG and of the filaments around it. For this task we put a slit along the BCG major axis; this is shown as a rectangle in Figure 1,  which includes an unsharp-mask image of the BCG, created by subtracting a smoothed version from the $r^\prime$ image. The integrated BCG spectrum is shown in Fig.~\ref{espectros}. In order to obtain some spatial resolution from the BCG slit spectrum, we extracted spectra from each pixel along the slit direction. The slit comprises 110 pixels and after binning we extracted 66 spectra along it, with roughly the same S/N $\sim$ 30 (as measured in the wavelength interval $6300 - 6500$~\AA). 
The BCG emission lines main parameters (central positions, fluxes and FWHMs) have been extracted using {\tt IRAF SPLOT}. For the H$\alpha$ + [NII] doublet line set and the [SII] doublet we performed a simultaneous fit assuming Gaussian profiles for the lines.
      
Galaxy spectra is affected by dust in our Milky Way and in the galaxy in consideration. The extinction due to our galaxy was corrected using CTIO standard extinction function. For the analysis in Sect. 4, where we will discuss diagnostic diagrams, it is necessary additionally to correct line emission for the reddening due to dust in the galaxy itself. Although some line ratios are insensitive to reddening (as those in the BPT diagram), this is not the case for all line ratios discussed in this work. We have estimated the galaxy reddening from the observed ratio of H$\alpha$ and H$\beta$ lines. We have adopted the \citet{Fitzpatrick99} extinction law, assuming $R_V=3.1$ and an intrisic line ratio equal to 2.86, appropriate for a low-density gas with temperature of $10^4$ K and assuming the recombination case B  \citep{Oster89}. We measured the observed H$\alpha/$H$\beta$ ratio for the total spectra and for each individual spectrum and we obtained a ratio of $3.57 \pm 0.17 $ and $3.61 \pm 0.08$ for the total spectrum and for the mean of the segmented regions, respectively. These values are consistent with a similar analysis performed by \cite{Ogrean10}.

%<--Figure #4

\subsubsection{Identification of a double gravitational image}
\label{sec.grav.arc}

\begin{figure}
\begin{center}
\includegraphics[width=0.8 \columnwidth,angle=270]{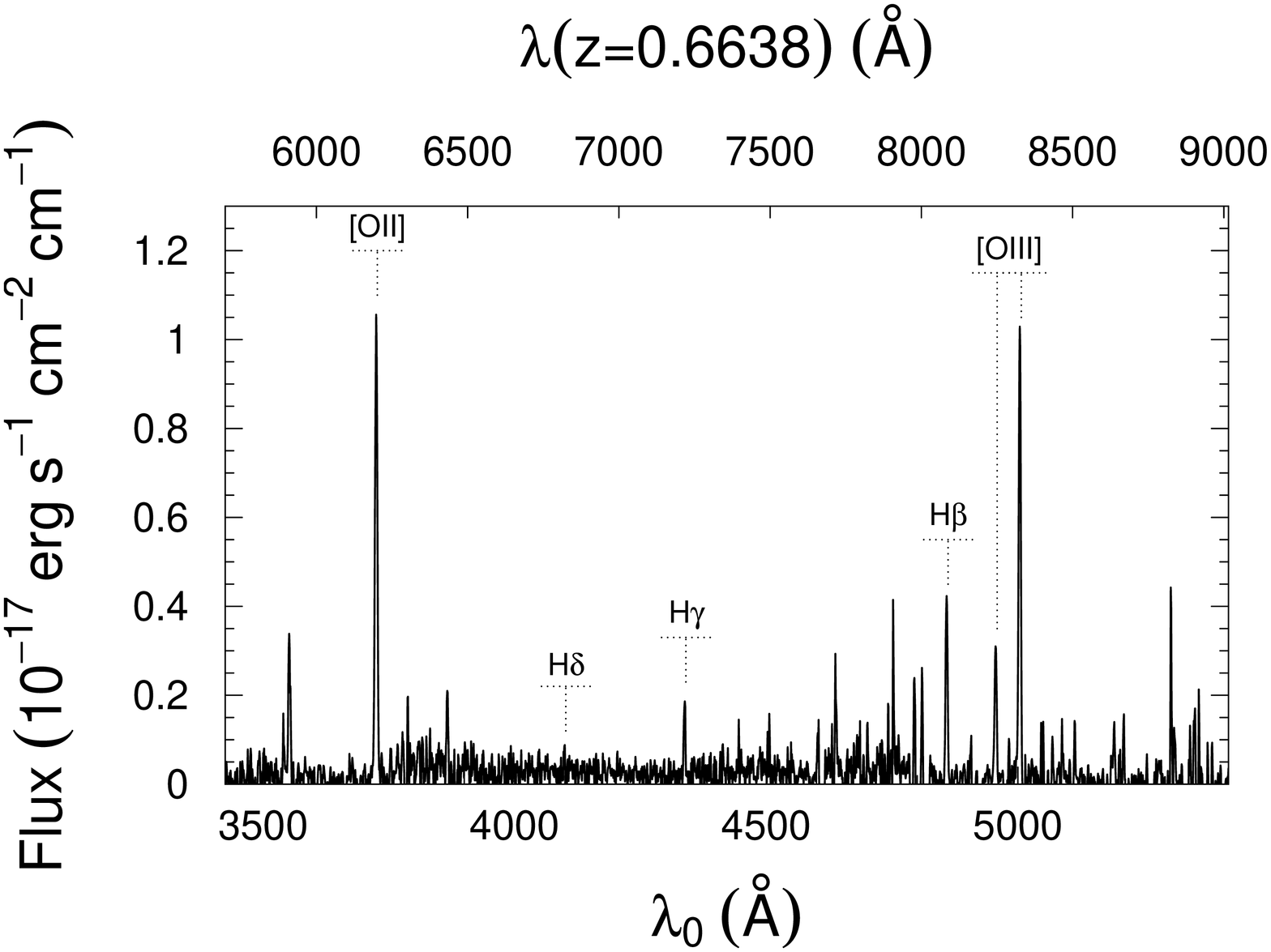}\quad
\includegraphics[width=0.8 \columnwidth,angle=270]{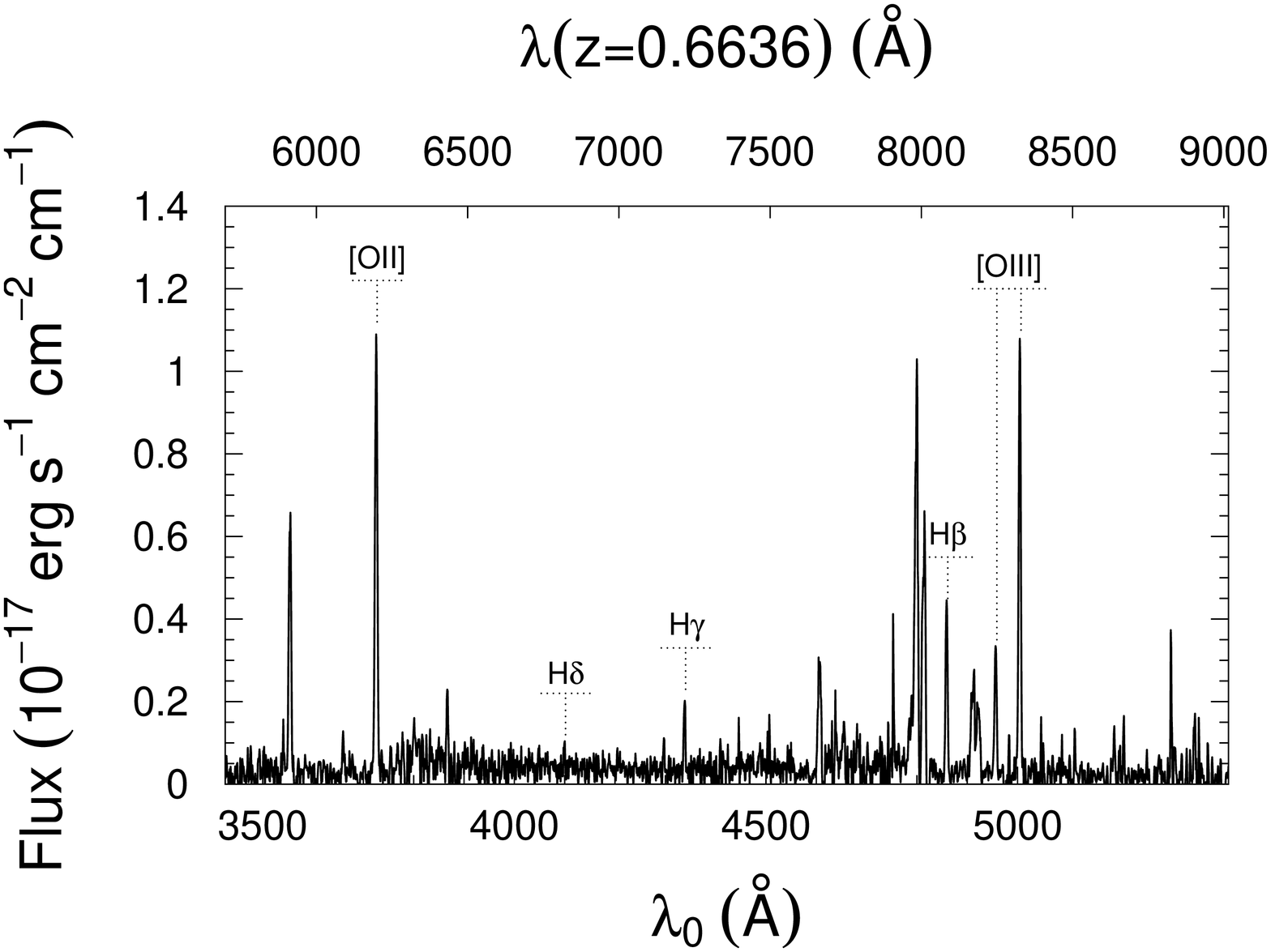}\quad
\caption{Spectra of the arcs A (top) \& B (bottom). Whereas arc~A appears in the field (see Fig.~\ref{cortes}) as an arclike structure with a bright knot, arc~B can not be directly seen due its proximity with the bright BGC. The measured redshifts are $z_A=0.6638 \pm 0.0004$ and $z_B=0.6636 \pm 0.0004$.}
\label{strong_lens}
\end{center}
\end{figure}

%<--Discovery of the arcs
A double gravitational arc (hereafter A \& B) was discovered through the BCG spectrum, when we noted pairs of lines above and below the main continua (Fig.~\ref{espectros}), corresponding to the same redshift of $z=0.664$, as we can see in Fig.~\ref{strong_lens}.

%<--Arcs features
Arc~A can be seen in the image (Fig.~\ref{cortes}), as well as its continua on the spectrum. On the other side, arc~B is probably fainter and its image is projected over a much brighter region of the BCG, thus we could detect only its emission line system. The arcs A and B are nearly perfectly aligned with the BCG optical centre and are, respectively, $4.80$ arcmin and $5.22$ arcmin away from it.

%<--Figure #5

\section{The Mass of RX1504}
\label{sec_mass}
In this section we present two distinct mass estimates of the cluster, the first adopting virial scaling relations and the second based on a joint analysis of strong and weak lensing.

\subsection{The velocity distribution}
\label{ssec_velocity}

\begin{figure}
\begin{center}
\includegraphics[width=0.35\textwidth,angle=270]{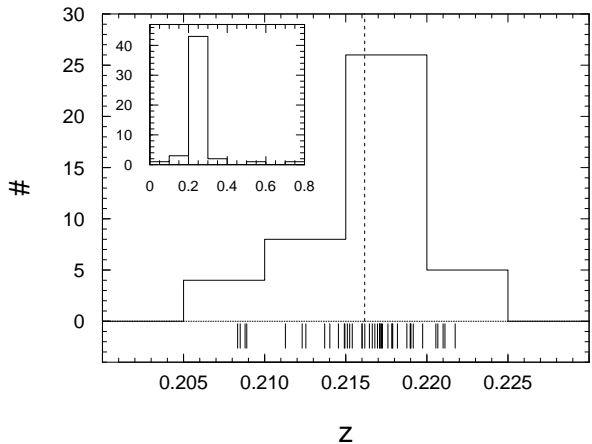}
\caption{Redshift distribution of galaxies in the field of RXC J1504-0248. The smaller panel shows the distribution of all galaxies while the bigger panel shows only the cluster member galaxies. The dotted line shows the mean.}
\label{zhist}
\end{center}
\end{figure}

Fig.~\ref{zhist} shows the redshift distribution obtained with the data in Table \ref{tabelaxcsao}. From the 174 galaxies brighter than $r' = 22.15$, we have obtained  42 reliable redshifts, leading to a completeness of 27\%. We consider as cluster members the 32 galaxies with spectroscopic redshifts in the interval $0.209 < z < 0.222$ (or $62700 < v <  66480$ km s$^{-1}$). According to the Anderson-Darling test, the null hypothesis of a Gaussian distribution cannot be ruled out within 99\% c.l. (${\rm p-value}=0.22$). The cluster mean radial velocity is ${cz} = 64878 \pm 176$ km s$^{-1}$ (or $z = 0.2164 \pm 0.0006$) and its radial velocity dispersion is $\sigma_{v}/(1+z) = 794_{-83}^{+123} $ km s$^{-1}$.

%<--Figure #6

From this data it is possible to obtain a virial mass estimate for RXC J1504-0248. We have used two different virial scaling relations, presented by \cite{Biviano06b} and  \cite{Evrard08}, in order to check its consistency. We found very close values $M_{200} = 5.8\pm1.9\times 10^{14}$ $h^{-1}_{70}$ M$_\odot$ and $M_{200} = 5.9\pm1.9\times 10^{14}$ $h^{-1}_{70}$ M$_\odot$, respectively, for each scaling relation.  We have computed the same virial radius of the system as $R_{200}=1.6\pm0.2$ $h_{70}^{-1}$ Mpc. In the NED database  there are 13 additional galaxies with measured radial velocities and that can be considered cluster members \citep[][]{Eisenstein11,Rines16}. They are also shown in Table \ref{tabelaxcsao}. Adding these new data, we obtain for both scaling relations $M_{200} = 6.8 \pm 1.9 \times 10^{14}$ $h^{-1} _{70}$ M$_\odot$ and $R_{200}= 1.7 \pm 0.2$ $h_{70}^{-1}$ Mpc, in good agreement with our previous estimates.

\subsection{Gravitational lensing}

\label{ssec_lensing}
We now discuss the cluster mass estimate based on a joint weak lensing plus strong lensing analysis. Here we will just introduce the basic concepts as used in our analysis. For more details on gravitational lensing  we refer the reader to one of the  reviews available in the literature \citep[e.g.][]{Meylan06}.

\subsubsection{Basic concepts}

%<--About gravitational lensing 
The gravitational lensing  technique is related to the measurement of the distortion caused in the light path by a large amount of mass. This phenomenon may lead a source (e.g. a background galaxy) to have its apparent position and image shape changed. By measuring the deformation and/or the position of the images, we can recover the mass that acted as a lens. The lensing can be described in terms of the  convergence, 
\begin{equation}
\kappa = \frac{\Sigma}{\Sigma_{cr}}\,,
\label{eq:kappa}   
\end{equation}
defined as the ratio of the projected surface mass density of the lens $\Sigma$, and the lensing critical density
\begin{equation}
\Sigma_{cr} = \frac{c^2}{4\pi G}\frac{D_s}{D_{ds} D_d}\,,
\label{eq:Sigmacr}   
\end{equation}
where $D_s$, $D_{ds}$ and $D_d$ are the angular diameter distances to the source, between the lens and the source, and to the lens respectively.

%<--Define the regimes
The strong regime occurs when $\kappa \geq 1$ and corresponds to the most dramatic effects, when a background galaxy can be seen as a multiple image depending on its relative position in relation do the lens axis (for the spherical simmetry in the mass distribution assumed here). On the other side, the weak regime corresponds to $\kappa\ll 1$, when the lens effect is almost linear and can be described as a function of the  convergence $\kappa$ and an anisotropic distortion represented by the shear, $\gamma=\gamma_1+i\gamma_2$, where
\begin{equation}
\gamma_1 = |\gamma| \cos(2\theta)\, 
\quad\mbox{,}\quad
\gamma_2 = |\gamma| \sin(2\theta) \,  
\label{eq:gamma.def} 
\end{equation}
and $\theta$ is the shear orientation on the plane of the sky. Both $\kappa$ and $\gamma$  are related to the second derivatives of the 2D-projected gravitational potential.

%<--Observational quantities
Unlike the strong regime, the weak lensing effect can not be identified in a single galaxy. Since the lens effect is tenuous, it is necessary a large sample to assess the coherent ellipticity $\langle e \rangle$ induced by the lens, which is directly related to the effective shear $g$,
\begin{equation}
\langle e \rangle \simeq g \equiv \frac{\gamma}{1-\kappa}\mbox{.}
\end{equation}
The effective shear is represented by a spin-2 tensor (as well as $\gamma$ and $\langle e \rangle$). We  can represent its components  following the tangential direction in relation to the cluster centre, e.g. $g_+$, and another one $45^\circ$ in relation to that, $g_\times$.

\subsubsection{Analysis}

Our lens model assumes a circularly symmetric projected mass distribution described by a Navarro-Frenk-White profile \citep[NFW;][]{NFW} with the centre matching the BCG. The two main parameters of this model are the mass and the concentration, $M_{200}$ and $c$.

For the strong lensing analysis we consider as observable the gravitational arc position, assuming that the distance between the two multiple images is twice the Einstein radius $\theta_E$. This corresponds to the radius where %$\bar{\kappa}(\theta_E)=1$,
\begin{equation}
\bar{\kappa}(\theta_E)=1
\label{eq:kappa.bar}
\end{equation}
being $\bar{\kappa}$ the mean convergence inside a given radius \citep[e.g.][]{Wright00}.

\begin{figure*}
\begin{center}
\includegraphics[width=0.7\columnwidth, angle=-90]{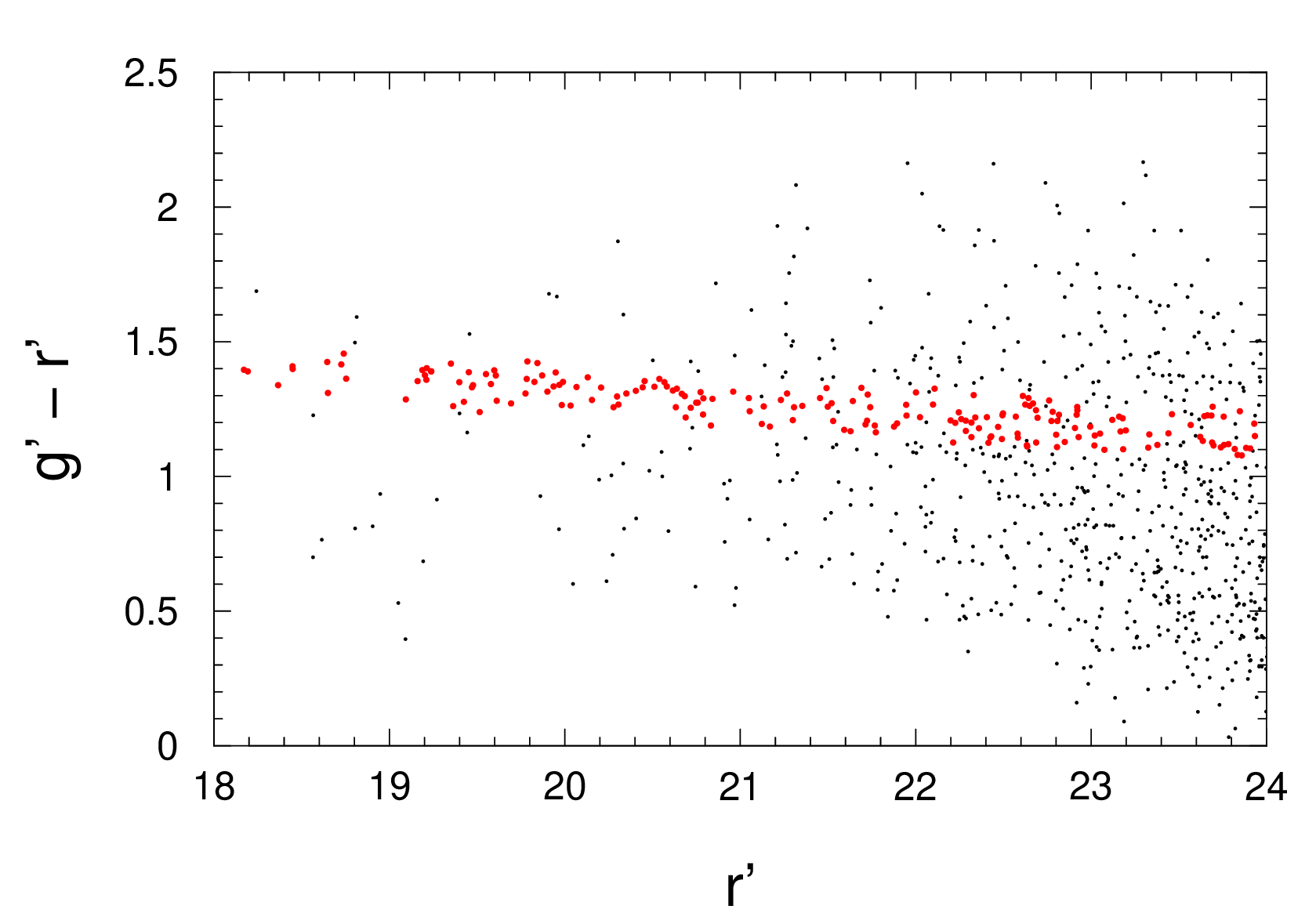}
\includegraphics[width=0.7\columnwidth, angle=-90]{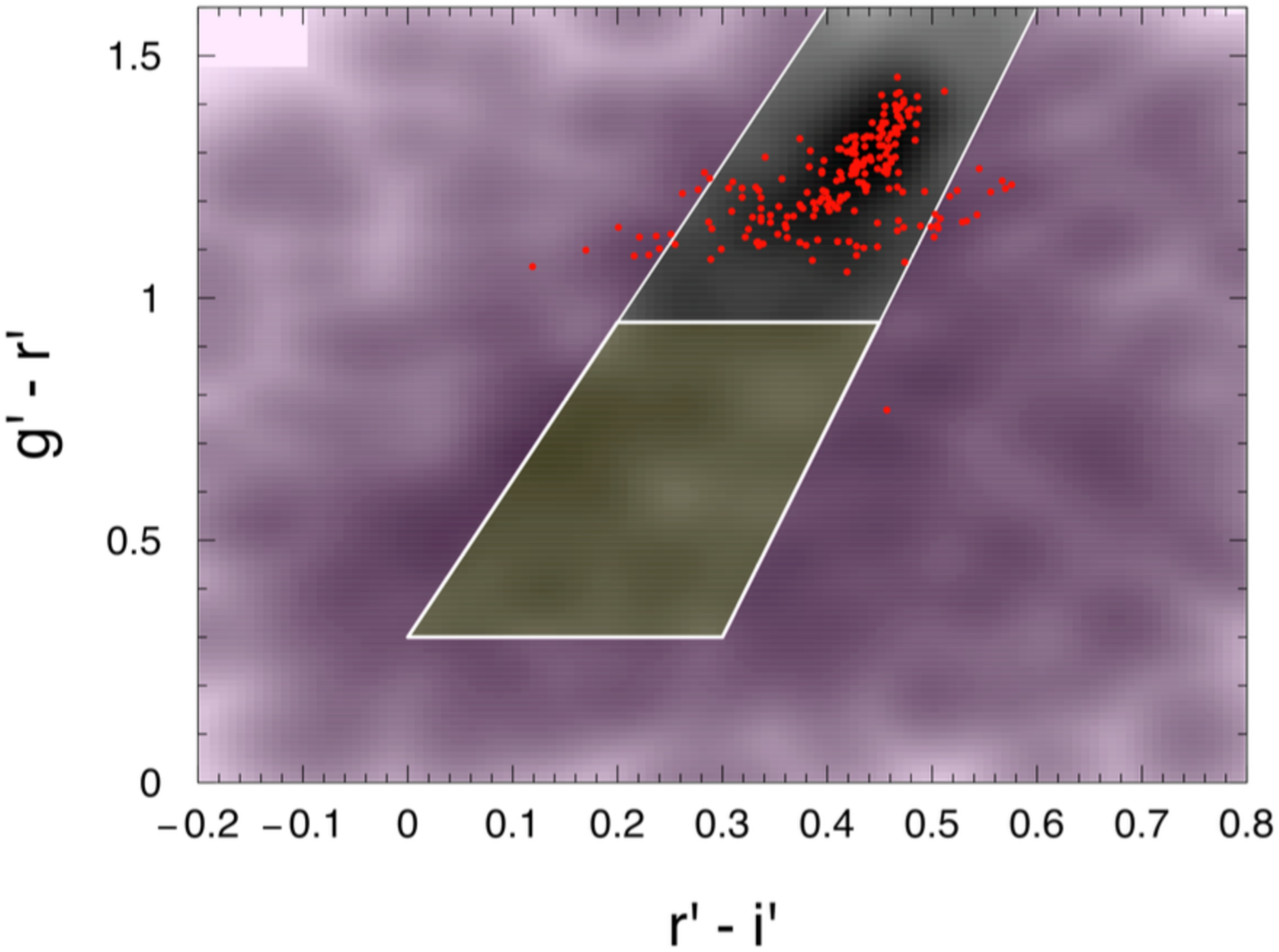} 
\caption{{\it Left:} Colour-magnitude diagram of our galaxy sample. We assume that the cluster members occupy a  stripe in this diagram known as  the red sequence (red points). {\it Right:} Galaxy colour-colour diagram. Using the CFHTLS galaxies as guidelines, we have identified the {\it locus} occupied by the foreground galaxies  (central polygonal region), i.e. those with smaller redshift than the cluster members. The red sequence cluster members correspond to the red points. Excluding these two regions, we found the {\it locus} where the background galaxies prevail (i.e. galaxies with larger redshift, magenta region).}
\label{fig:diagrams}
\end{center}
\end{figure*}

The weak lensing analysis depends essentially on the statistical measurement of the distortion of the images of galaxies in the cluster background.  Since we do not have spectroscopic redshifts for all galaxies in the field, we have used colour-magnitude  and colour-colour  diagrams to classify the observed galaxies in cluster members, background and foreground galaxies \citep[e.g.][]{Edu05,Zitrin12,Hoekstra13,MonteiroOliveira17a,MonteiroOliveira17b}. The red cluster members were found through the identification of their sequence in the colour-magnitude diagram (Fig.~\ref{fig:diagrams}). The identification of (candidate) foreground and background galaxies was performed through a joint analysis of our sample with the data collected by the CFHTLS T0004 Deep and Wide fields \citep{Coupon09}, which comprise observations of 35 square degrees in the  $u^*$$g^\prime$$r^\prime$$i^\prime$$z^\prime$ bands, providing catalogues of galaxies with accurate photometric redshifts. We constructed  colour-colour diagrams for galaxies in our sample and in the CFHTLS sample, considering only objects brighter than the limiting magnitude of each band (see Section \ref{ssec_imaging}). We then used CFHTLS galaxies to identify colour regions occupied mostly by background or foreground galaxies, and classified accordingly all galaxies in our images as we can see in Fig.~\ref{fig:diagrams}.

%<--Figure #7

The amount of distortion suffered by the image of a background galaxy depends on its angular distance to the cluster  and on its own redshift through the critical density (Eq.~\ref{eq:Sigmacr}).  We have estimated this quantity from the CFHT sample, applying the same colour and magnitude cuts as done in our sample, obtaining a  critical surface mass density of $\Sigma_{\rm cr}=3.5\times10^9$ M$_{\odot}$ kpc$^{-2}$, considering objects at the mean redshift of the background galaxies ($\bar{z}_{\rm back}= 0.80\pm 0.43$). 
   
To analyze the shape of the background objects we have used the software {\sc im2shape}, developed by \cite{Bridle98}\footnote{\texttt{htpp://www.sarahbridle.net/im2shape/}}. This program models each galaxy as a sum of two Gaussians\footnote{The stars are modelled as a single Gaussian.} with elliptical bases and convolved with the seeing, which is also modelled having a Gaussian profile. To obtain the best parameters, the software uses a Markov Chain Monte Carlo (MCMC) method to minimize the residuals between image and model. As stars can be considered point sources, their observed circularised images are due to  atmospheric and instrumental effects, described by the point spread function (PSF). We mapped the PSF (elipticity and size) variation along the field using 42 bona-fide stars. The shape of every galaxy was then deconvolved by a local PSF obtained by interpolation of the PFS of these stars.

Our lens model is designed to take into account both strong and weak lens observables to constrain the mass and concentration estimates, with different likelihoods for each dataset. For the weak lensing analysis (WL)  we assume that \citep[e.g.][]{MonteiroOliveira17a,MonteiroOliveira17b}.
\begin{equation}
\ln \mathcal{L}_{\rm WL}=-\dfrac{1}{2}\sum_{i=1}^{N_{{\rm b}}}   \frac{[g_{+,i}(M_{200},c)-e_{+,i}]^2}{\sigma_{\rm int}^2+\sigma_{i}^2}\mbox{,}
\label{likelihood.wl}
\end{equation}
where $N_{{\rm b}}$ is the number of background galaxies, $g_{+,i}$ is the modelled tangential shear, $e_{+,i}$ is the observed tangential ellipticity, $\sigma_{\rm int}$ is the intrinsic error (assumed as 25\%) and $\sigma_i$ is the measurement uncertainty \cite[see also][for more details on a similar weak lensing analysis]{Umetsu16}. 

The strong lensing likelihood (SL) is written as
\begin{equation}
\ln \mathcal{L}_{\rm SL}=-\dfrac{1}{2}   \left[\frac{\theta_E (M_{200},c)-\bar{\theta}_E}{\sigma_{\bar{\theta}_E}}\right]^2\mbox{,}
\label{likelihood.sl}
\end{equation}
where  $\bar{\theta}_E=5.01\pm0.50$ arcmin corresponds to the mean distance of the arcs to the BCG centre and  $\theta_E (M_{200},c)$ is the NFW modelled Einstein radius obtained through the numerical solution of Eq.~\ref{eq:kappa.bar}. For this purpose, we have applied the modified quasi-Newton method \citep[L-BFGS-B;][]{Byrd95} implemented on function {\sc optim} in the {\sc R} environment \citep{R}.

As weak and strong lensing models depends on independent data, the combined likelihood can be written as
\begin{equation}
\mathcal{L} = \mathcal{L}_{\rm WL} \mathcal{L}_{\rm SL}\mbox{.}
\label{likelihood.combined}
\end{equation}
Additionally, in the  Bayesian framework, the posterior probability of the parameters is given by
\begin{equation}
{\rm Pr}(\rm M,c| data) \propto  \mathcal{L}({\rm data}|M,c)\times\mathcal{P}(M,c)\mbox{,}
\label{eq:posterior}
\end{equation} 
where $\mathcal{P}(M,c)$ is the prior of the parameters.

We have mapped the posterior (Eq.~\ref{eq:posterior}) using the MCMC algorithm with a simple Metropolis sampler implemented on the {\sc R} package {\sc MCMCmetrop1R} \citep{Martin11}. We have generated a chain with  $1\times10^5$ elements, large enough to ensure convergence, and  neglecting the first $5\times10^4$ samples as ``burn-in'' steps. Additionally, we have adopted non-informative uniform priors for $M_{200}$, $0   \leq M_{200} \leq 1 \times 10^{16}$ M$_\odot$ and $c$, $0.1\leq c \leq 40$, in order to accelerate the model convergence \citep[e.g][]{MonteiroOliveira17a,MonteiroOliveira17b}. Aiming to check the consistency of our datasets, we have estimated the posterior, Eq.~\ref{eq:posterior}, also for the individual likelihoods given by Eqs.~\ref{likelihood.wl} and \ref{likelihood.sl}. We have also considered cases where additional constraints were added to the NFW concentration parameter, $c$, which, as we will see, is not well constrained by our data. The best estimated parameters for several situations are presented in Table ~\ref{tab:lens}. 

%<--Lensing results
\begin{table}
\begin{center}
\caption[]{Gravitational lensing analysis results. SL(1) and SL(2) correspond to the dual solution for the measured arc positions (strong lensing only). WL$^{\ddagger \ddagger}$ shows the best solution for the weak lensing data set alone, keeping both mass and concentration as free parameters. WL$^{\ddagger}$ corresponds to the same previous data set, but the concentration was set by the \cite{Duffy08} relation. WL+SL$^{\dagger \dagger}$  shows the parameters obtained for the joint weak and strong analysis when we kept both mass and concentration as free parameters. Finally,  WL+SL$^{\dagger}$ corresponds to the joint analysis for a fixed concentration based on SL(1) results.}
\begin{tabular}{lcc}
\hline
\hline
Dataset & M$_{200}$ ($10^{14}$~M$_\odot$) & NFW concentration \\
\hline
SL(1)         & $5.8_{-0.4}^{+0.7}$ &  $6.4_{-0.6}^{+0.5}$  (free)\\[5pt]
SL(2)         & $2.6_{-0.6}^{+0.4}$ &  $9.4_{-0.3}^{+0.6}$  (free)\\[5pt]
WL$^{\ddagger \ddagger}$ & $26_{-24}^{+49}$    &  $1.4_{-1.2}^{+3.4}$
(free)\\[5pt]
WL$^{\ddagger}$     & $7.8_{-2.7}^{+2.3}$ &  $3.3$
(M-c relation)\\[5pt]
WL+SL$^{\dagger  \dagger}$     & $2.0_{-1.1}^{+0.8}$ &  $10.3_{-2.9}^{+2.2}$  (free) \\[5pt]
WL+SL$^{\dagger}$  & $5.3\pm0.4$ &  $6.4$ (M-c relation)\\ 
\hline
\hline
\end{tabular}
\label{tab:lens}
\end{center}
\end{table}

\begin{figure}
\begin{center}
\includegraphics[angle=90, width=\columnwidth]{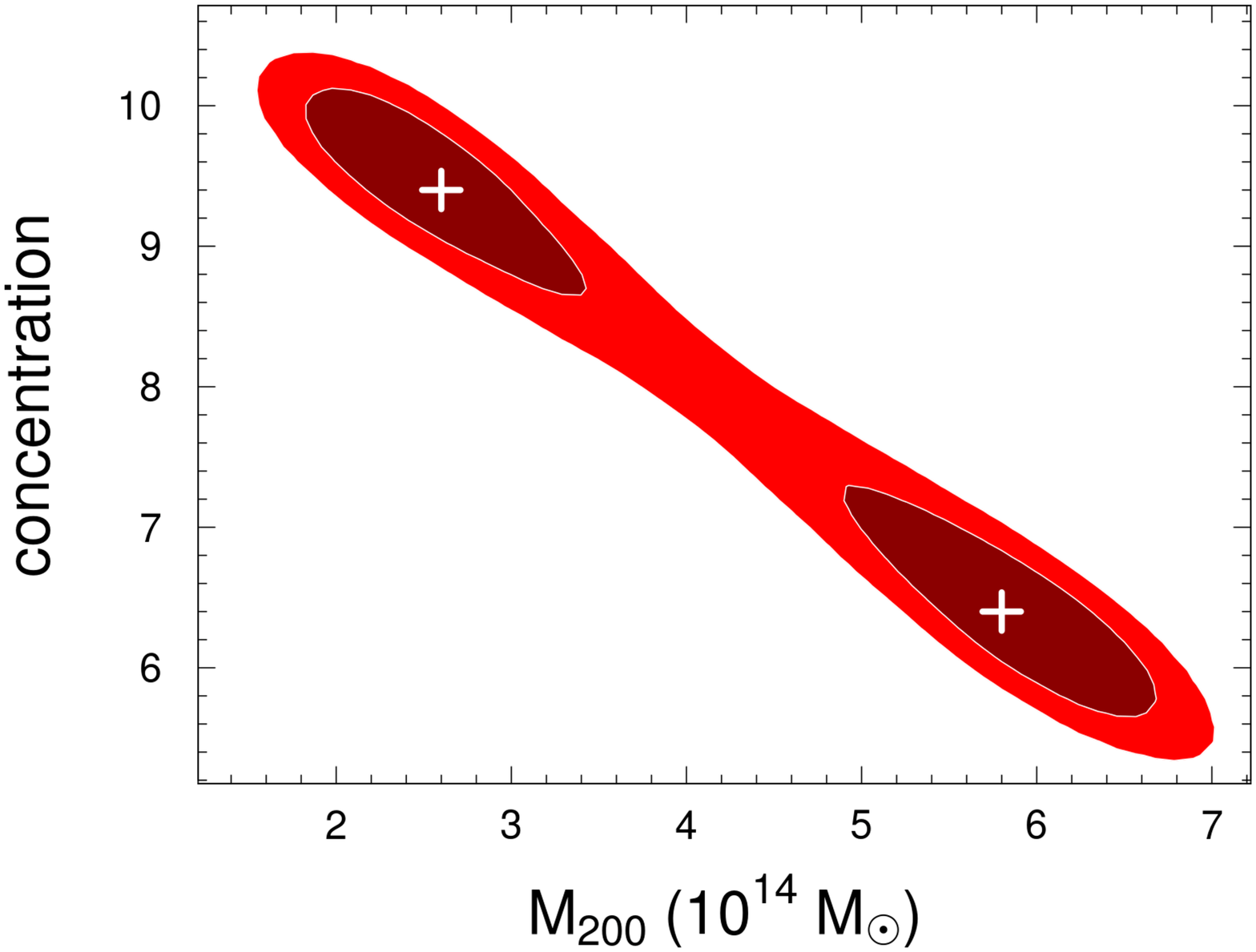}  
\caption{Posterior of $M_{200}$ and concentration as revealed by the analysis considering only the strong lensing  dataset (Eq.~\ref{likelihood.sl}). The dotted and straight lines correspond, respectively to 68\% and 95\% c.l. The $M_{200}-c$ relation leads to two parirs of solutions for Eq.~\ref{eq:kappa.bar}, highlighted as ``+": M$_{200}=5.8_{-0.4}^{+0.7}\times 10^{14}$ $M_\odot$ with $c=6.4_{-0.6}^{+0.5}$,  and $M_{200}=2.6_{-0.6}^{+0.4}\times 10^{14}$ M$_\odot$ with $c=9.4_{-0.3}^{+0.6}$, respectively SL(1) and SL(2).}
\label{fig:strong.res}
\end{center}
\end{figure}

%<--Individual dataset results
We start this analysis by considering only the strong lensing dataset, whose likelihood is given by Eq.~\ref{likelihood.sl}. The posterior shows a bimodal $M_{200}-c$ relation, as seen in Fig.~\ref{fig:strong.res}, with solutions, $M_{200}=5.8_{-0.4}^{+0.7}\times 10^{14}$ M$_\odot$ with  $c=6.4_{-0.6}^{+0.5}$ and $M_{200}=2.6_{-0.6}^{+0.4}\times 10^{14}$ M$_\odot$ with $c=9.4_{-0.3}^{+0.6}$, hereafter SL(1) and SL(2), respectively. On the other side, the weak lensing dataset alone has proved unfruitful to constrain simultaneously $M_{200}$ and $c$, returning estimations with large error bars ($M_{200}^{WL^{\ddagger \ddagger}}=26_{-24}^{+49}\times 10^{14}$ M$_\odot$ with  $c^{WL^{\ddagger \ddagger}}=1.4^{+3.4}_{-1.2}$). Moreover, the MCMC algorithm reported a non-stable behaviour of the chain (characterised by the  trace of the variable) contributing to make this results untrustful. For the sake of comparison, we computed the weak lensing mass keeping the concentration fixed by the $M_{200}-c$ relation presented by \cite{Duffy08}, finding $M_{200}^{WL^{\ddagger}}=7.8_{-2.7}^{+2.3}\times 10^{14}$ M$_\odot$ with $c^{WL^{\ddagger}}=3.3$.

%<--Figure #8

%<--Combined dataset results
The mass estimated from the combined WL+SL$^{\dagger \dagger}$ analysis, M$_{200}^{\rm WL+SL^{\dagger \dagger}}=2.0_{-1.1}^{+0.8}\times 10^{14}$~M$\odot$, is very close to the SL(2) case and clearly dominated by the SL data. It is also significantly lower than those estimated in Sec.~\ref{ssec_velocity}.
We tend to disfavour this solution in favour of the other given the large disagreement with our previous estimation. This bimodality is probably caused by the relatively poorness of out data  and the simplicity of our model that, for instance, assumes  circular symmetry. To overcome this,  we have  proceeded the joint modelling WL+SL$^{\ddagger}$ with a fixed concentration given by the solution SL(1). From this procedure we have obtained $M_{200}=5.3\pm0.4\times10^{14}$ M$_\odot$ within $R_{200}=1.56\pm0.04$ $h^{-1}$ Mpc.

%<--Comments about the lensing results
This mass estimate (as well as $R_{200}$) is in excellent agreement with the virial mass obtained in Section \ref{ssec_velocity} ($5.9 \pm 1.9 \times 10^{14}$ $h^{-1}$ M$\odot$) and is consistent with the two \cite{Zhang12} results: the dynamical mass, $M_{200}=6.43 \pm 0.65 \times 10^{14}$ $h^{-1}$ M$_{\odot}$, obtained through spectroscopic data from VIMOS for 53 member galaxies, and the hydrostatic mass,  $M_{200}=7.65 \pm 0.65 \times 10^{14}$ $h^{-1}$ M$_{\odot}$, obtained from XMM-{\it Newton} X-ray data. The consistency between all these mass estimators (hydrostatic masses assume equilibrium, whereas lensing masses do not) may be considered an indication that RX1504 is in a state of overall dynamical equilibrium.

All these mass estimates indicate that RX 1504 is a massive cluster. Indeed, using, e.g., the Schechter cluster mass function of \cite{Girardi98} for nearby clusters, it can be shown that it is among the $21\%$ most massive clusters with masses above the characteristic mass $M^* = 2.6_{-0.6}^{+0.8} \times 10^{14}$ $h^{-1} M\odot$.

\section{The emission by the filamentary gas}
\label{sec_gas}   
The spectroscopic data described in Section \ref{sec_data} can help to understand the physical conditions of the gas and provide clues on its ionization mechanism. In this section we discuss the line emission coming from the filamentary structure associated to the BCG of RX1504.

\subsection{Morphology and kinematics of the filamentary structure}
   \label{ssec_morphology}
In order to highlight the filamentary features in the central galaxy, we adopted an unsharp masking technique. To this end, we produced two smoothed images, the first one convolved with a 1 pixel (0.146$^{\prime\prime}$) 2D Gaussian kernel to filter out the high frequency noise and the second one is the original image smoothed with a 3 pixel kernel. The width of this kernel was fine-tuned to increase the visibility of the filaments, with larger kernels washing away the filament in the final image. The unsharp masked image is obtained simply by subtracting the first image by the second (the mask). The result is that the low frequency (large scale) features of the original image are filtered out, and point-like structures as well as narrow, sharp filaments, are accentuated. This procedure was done with the 3 bands and a composite $g'r'i'$ image is shown at the bottom panel of Figura \ref{cortes}. 
  
A careful analysis of this figure reveals a bright filament which extends by $\sim 38.4$ $h^{-1}_{70}$ kpc from north-east (NE) to south-west (SW) (the position angle of the NE extension is $\sim 55^\circ$), crossing or associated to a prominent emission in the central region of the galaxy. This morphology is also apparent in the UV from HST observations \citep{Tremblay15}, as well as in optical line emission
measured with VIMOS in IFU mode \citep{Ogrean10}. The BCG lies at the  peak of the X-ray emission, which is also aligned with the filament \citep[see Fig. 9 of ][]{Bohringer05}.

\begin{figure}
\begin{center}
\includegraphics[width=1.05 \columnwidth,angle=0]{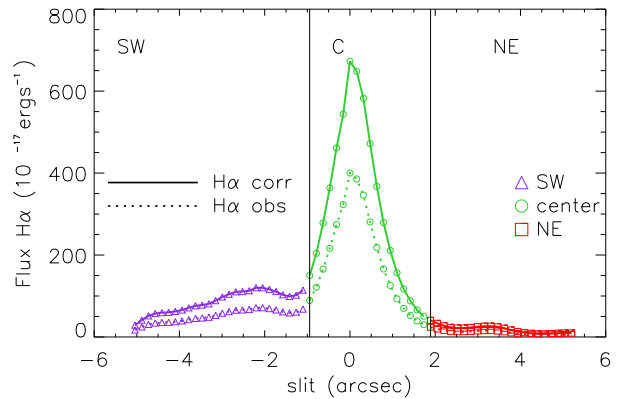}
\caption{Flux of the H$\alpha$ emission line (observed and extinction corrected) along the slit. The slit is divided in three regions, corresponding to the SW, central and NE emission.}
\label{perfil_ha}
\end{center}
\end{figure}

Fig.~\ref{perfil_ha} shows the profile of the H$\alpha$ emission line along the slit. To help in the discussion below, we found useful to divide the filamentary emission in three regions, corresponding to the SW, central (1 arcsec of each side of the center), and NE regions. The line emission peaks in the centre and is stronger in the SW than at the NE region. It can be shown that this overall behaviour is shared by other emission lines and by the $r^\prime$ filamentary emission.

%<--Figure #9

\begin{figure}
\begin{center}
\includegraphics[width=1.05 \columnwidth,angle=0]{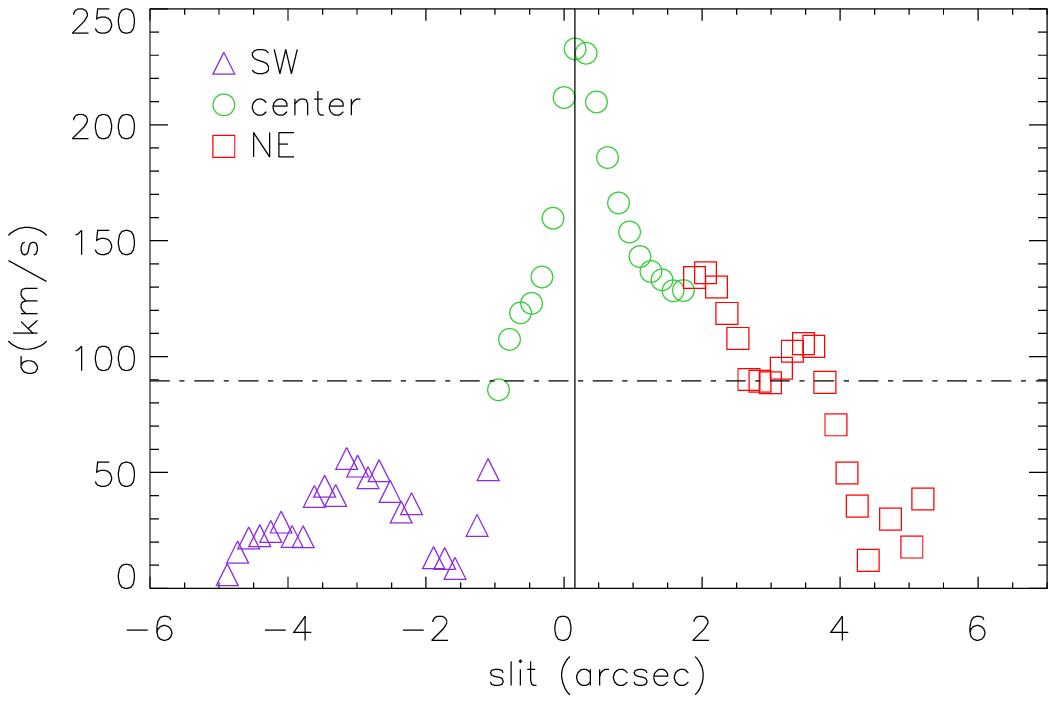}\\
\includegraphics[width=1.05 \columnwidth,angle=0]{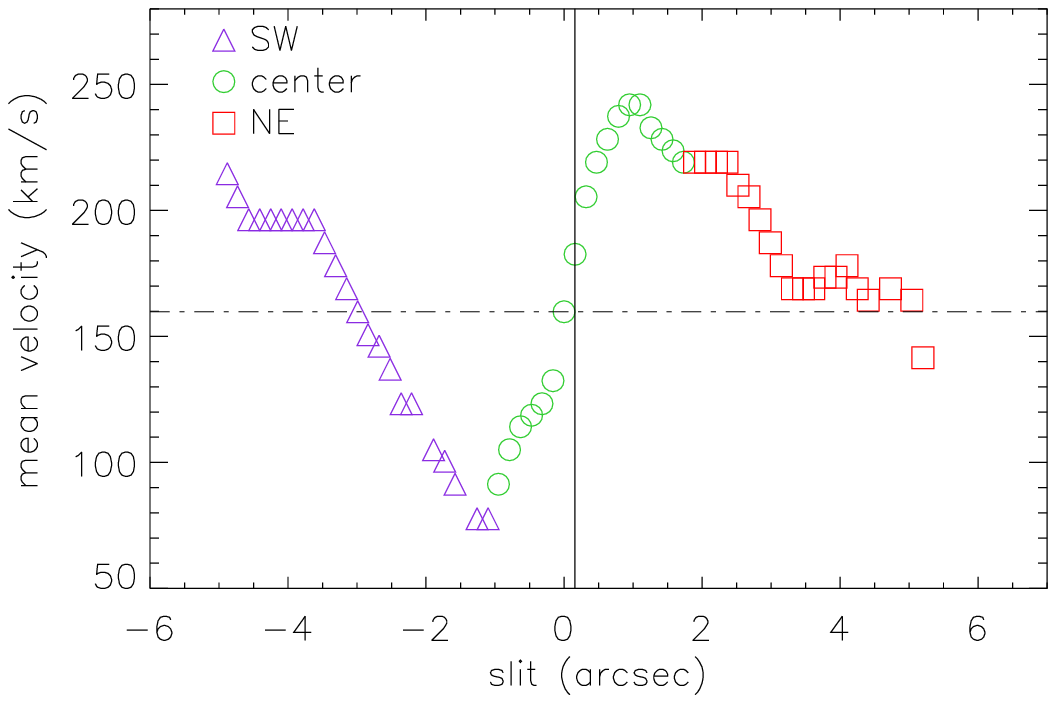}
\caption{ Top: velocity dispersion of the H$\alpha$ emission line along the slit. The dashed and solid lines show the mean velocity and the maximum velocity position, respectively. Bottom: mean velocity of the H$\alpha$ emission line along the slit. The solid line shows the same position of the top panel.}
\label{kinematics}
\end{center}
\end{figure}

We can also obtain kinematical information from the slit spectrum. Fig.~\ref{kinematics} (top) shows the velocity dispersion $\sigma$ along the slit, measured from the H$\alpha$ line, here defined as $\sigma = \sqrt{\sigma_{obs}^2 - \sigma_{inst}^2}$, where $\sigma_{obs}$ and $\sigma_{inst}$ are the observed and the instrumental width, respectively. The velocity dispersion is, as expected, strongly peaked in the galaxy centre, $\sigma \sim 250$ km s$^{-1}$ and, after $\sim 12$ kpc, falls to $\sim 50$  km s$^{-1}$ towards the SW and NE filamentary extensions.
Interestingly, the mean velocity along the slit (after subtracting the mean velocity in the galaxy centre) shows a pattern very similar to a rotation curve, with a bulge component of radius $\sim 4$ kpc and  velocity up to $\sim 70$ km s$^{-1}$, falling at large radii (Fig.~\ref{kinematics}, bottom). However, since we are working with slit spectrum (not an IFU) and that we do not know the actual geometry of the emission region, other interpretations are possible. For instance, the infall of the gas which feeds the galaxy center and the filaments may produce a pattern similar to the observed. 

%<--Figure #10
   
\subsection{Diagnostic diagrams}
\label{ssec_ionization}
\begin{figure}
\begin{center}
\includegraphics[width=0.5 \textwidth,angle=0]{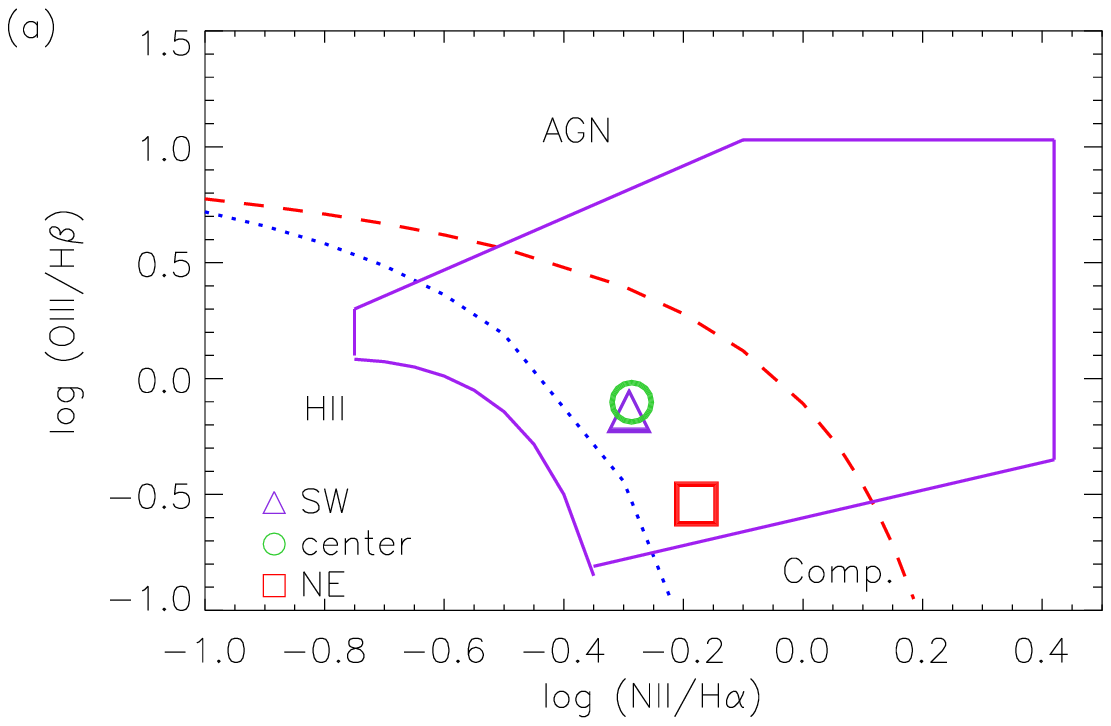}\quad
\includegraphics[width=0.5\textwidth]{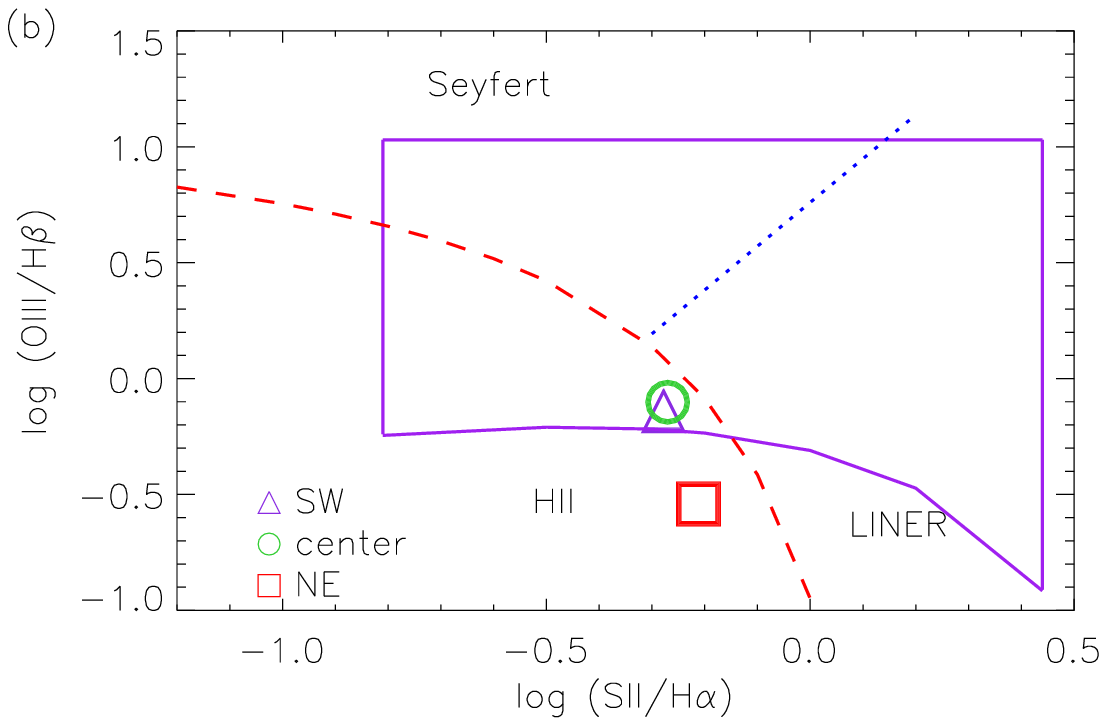}\quad
\includegraphics[width=0.5\textwidth]{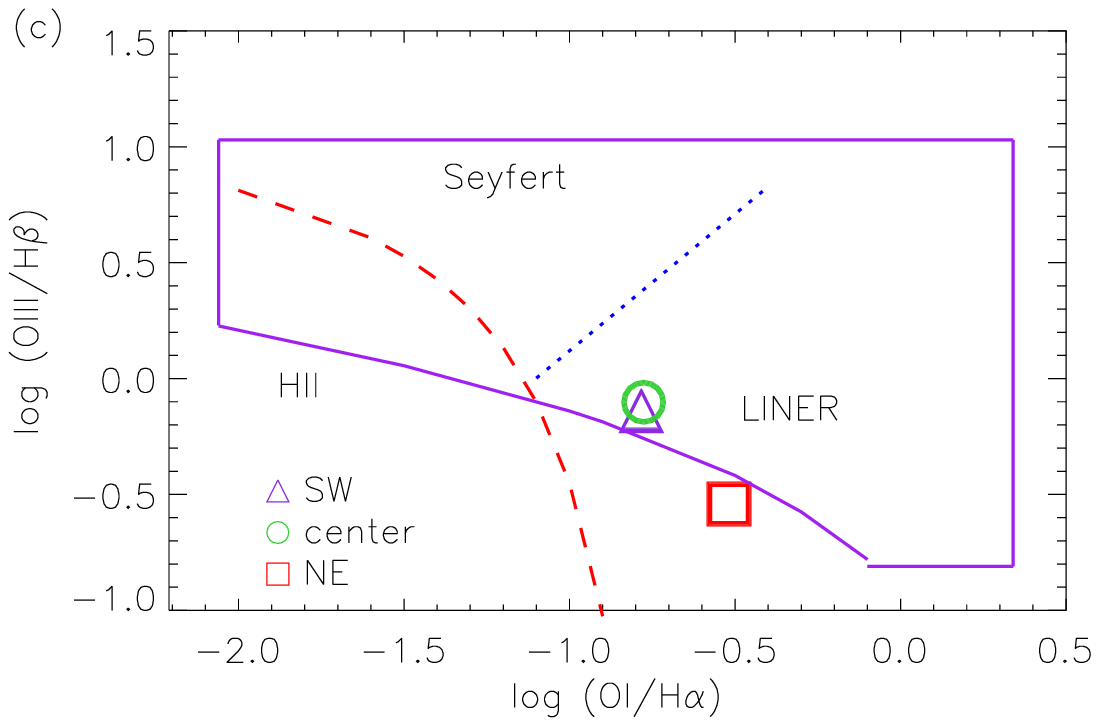}
\caption{The upper picture shows the BPT diagram with the limits set by \citet{Alatalo16} (polygonal region, straight purple line) for emission due to shock excitation, the empirical line from \citet{Kauffmann03} (dotted blue line)  and the theoretical limiting line proposed by \citet{Kewley01} (dashed red line). The two pictures under the BPT are alternative diagnostic diagrams with the same limits. The different symbols (triangle, circle,square) show the positions occupied by each region of the BCG in the diagrams (SW, Center and NW, respectively).}
\label{bpt_shocks}
\end{center}
\end{figure}

In this section we analyze the emission lines along the slit with the help of diagnostic diagrams, which were first proposed by  \citet{Baldwin81} to classify extragalactic sources according to their excitation mechanism. Fig.~\ref{bpt_shocks} (top) presents [O III]/H$\beta$ versus [N II]/H$\alpha$ (the traditional ``BPT'' diagram), the most used of the diagnostic diagrams\footnote{The actual lines used in the diagnostic diagrams discussed in this section are [O III]$\lambda 5007$, [O I]$\lambda 6300$, [N II]$\lambda 6584$ and [S II]$\lambda\lambda 6716,6731$, besides the Balmer lines H$\alpha$ and H$\beta$.}. When displayed with large numbers of objects \citep[][]{Kauffmann03, Mateus06}, this diagram shows two well-defined wings. The left wing contains essentially star-forming galaxies (SF), whereas the right wing is populated by galaxies with either Seyfert-like or LINER-like spectra. The SF wing appears due to the strong coupling in these galaxies between the O/H and N/O abundance ratios, the ionizing radiation field, and the ionization parameter. Objects outside this wing have stronger collisionaly excited lines (e.g., [O III], [N II]) indicative of ionization by a radiation field harder than that expected from massive stars \citep[e.g.][]{Cid10}. Fig.~\ref{bpt_shocks} also shows two lines. The dashed red line is an  empirical division proposed by \citet{Kauffmann03} to separate star-forming galaxies from AGNs. The dotted blue line is the theoretical boundary proposed by \citet{Kewley01}, corresponding to an upper envelope in which the ionization source is provided by young star clusters. Notice that, despite one or another ionization mechanism may dominate in each part of the BPT diagram, both stellar and non-stellar sources may be actually present. Indeed \cite{Grazyna06}  argue that the \cite{Kewley01} line corresponds to an AGN contribution of roughly 20\%. 

We will discuss below whether shocks in the filamentary gas can be a relevant source of excitation in RX1504.
If clouds formed during the cooling of the gas in the central part of the cluster are supersonic, their motion through a gaseous environment produce shocks, whose dissipation may be an important source of ionizing photons.
The region occupied by shocks in each diagnostic diagram, according to  \citet{Alatalo16}, is presented as a polygon delimited by straight purple lines in each panel of Fig.~\ref{bpt_shocks}. Interestingly, the line-ratios for the three regions of the BCG all fall into the shock region of Fig.~\ref{bpt_shocks}a, between the two lines mentioned above.

Besides the traditional BPT diagram, other diagnostic diagrams (also discussed in the original BPT paper) are also useful for our study, since they provide alternative diagnostics that allows checking our results. This is the case for the [OIII]/H$\beta$ versus [SII]/H$\alpha$ and the [OIII]/H$\beta$ versus [OI]/H$\alpha$ diagrams. They are considered less efficient than the usual BPT diagram for classification of galaxy emission, both because the SF/AGN dichotomy is not so clear and because simple photoionization models underpredict the [SII]/H$\alpha$ and [OI]/H$\alpha$ ratios \citep{Grazyna06}. Nevertheless, they are useful to complement this analysis since the consistency of the analysis with these three diagrams may be an evidence of the robustness of our results. Figs.~\ref{bpt_shocks}b and \ref{bpt_shocks}c show these alternative diagrams. They also display the \citet{Kewley01}  limit and the \citet{Alatalo16} shock-regions. The dotted blue line represents a Seyfert/LINER division proposed by \cite{Kauffmann03}. We can see that the {\it loci} occupied by RXJ1504 regions are ambiguous with respect to SF or AGN origin, but almost all regions are located within the shock limits. Indeed, whereas the SW and central regions of the filamentary structure falls in to the shock region in the three diagnostic diagrams, the NE region falls out of it in the  [OIII]/H$\beta$ versus [OI]/H$\alpha$ diagram, but close to it anyway.

In next section we will discuss these results. Now we only want to point out that the position of the regions in these diagnostic diagrams, out of the SF region (left wing in the BPT diagram), indicates that a fraction of the line intensities are due to processes other than star formation and consequently estimates of the star-formation rate through line intensities provide, in this case, only upper limits. With this caveat in mind, we have estimated the star formation rate from H$\alpha$ emission following the \citet{Kennicutt94}
 \begin{equation}
  SFR (M_{\odot}~ {\rm year^{-1}}) = 7.9 \times 10^{-42} L(H{\alpha}) {\rm (erg~ s^{-1})},
  \label{eq_sfr}
 \end{equation}
finding 168 $\pm$ 19 M$_{\odot}$ year$^{-1}$. Our result is lower (but still consistent within the errors) than the SFR =  241 $\pm$ 92 M$_{\odot}$ year$^{-1}$  found by \cite{Ogrean10} using UV data from the Hubble Space Telescope. However, \cite{Ogrean10} state that their value is not consistent with the number of O and B stars present in the BCG, which implies a lower SFR. On the other side, \cite{Mittal15}, using a star formation model consisting of an old stellar population plus a series of young stellar components, obtained $67 \substack{+49 \\ -27}$ M$_{\odot}$ yr$^{-1}$, below (but consistent with) our lower limit. 

%<--Figure #11
   
\section{Discussion}
\label{sec_disc}
What are the physical mechanisms responsible for the excitation of the filamentary structure? We now revisit the several processes invoked in the literature, using the results presented in the previous sections to provide new constraints  on the heating mechanisms of this structure.

AGN feedback may have a significant impact on the environment of the galaxy where the associated massive black hole resides through the effects of radiation, winds and jets \citep{Fabian12}. In particular, could be the filament observed in RX 1504 a jet produced by a central AGN? We have seen that the nuclear emission is consistent with a LINER; additionally, an examination of the permitted lines do not show evidence of extended wings that could be produced by jets or outflows. The radio observations do not show any evidence of radio lobes that could be considered an evidence of a jet. We conclude that AGN feedback in the form of a jet can not be the origin of the filamentary structure observed in this cluster. This is also a drawback for another possible ionizing source, cosmic rays, at least accordingly with the works of \cite{Jacob16a} and \cite{Jacob16b}, which require that these particles  be accelerated in a relativistic jet that, through the  excitation of Alfvén waves, transfers the cosmic rays energy to the cooling gas. To test this model, they analysed a sample of 39 sources, including RX J1504, and concluded that this kind of ionization is a plausible heating source, but only in the clusters which do not host  a radio minihalo, as is the case of  RX J1504.
 
The integral field observations by \cite{Ogrean10} have shown that the bulk of the line emission comes from the filamentary structure. As can be seen in the three diagnostic  diagrams discussed in the previous section, the filament emission falls either in the transition region or in the LINER region of these diagrams. Consequently, it is fair to assume that, although  star-formation may be contributing to the line emission of the filament, as argued by \cite{Ogrean10} and \cite{Mittal15}, an additional source of ionization is required \citep[e.g.][]{Kewley01,Grazyna08}. Actually, \cite{Ogrean10} also argues that their estimate of number of ionizing stars are just enough to ionize and heat the filaments and the line ratios are consistent with additional heat sources. The position of the line ratios in our diagnostic diagrams requires the consideration of LINERS as an alternative or additional ionization source, which will be better discussed in the next section. 

Thermal conduction was one of the first mechanisms studied to explain the heating of the cooling gas in clusters  \citep[e.g.][]{Binney81, Tucker83}. In these early works, the thermal conduction hypothesis adopted irrealistic physical parameters and was discarded. But more recent studies \citep[e.g.][]{Zakamska03} have shown that it may be a relevant heating source for some clusters, specially those that do not have a strong radio galaxy in this centre. However, the presence of a mini-halo radio emission in RX J1504 decreases the likelihood that thermal conduction plays a major role on the heating and excitation of the filament.

We now discuss in more detail two other possible sources of ionizing photons: LINERs and shocks.

\subsection{LINERs}

LINERs (Low Ionization Narrow Emission-Line Regions) were introduced by \cite{Heckman80} to designate active nuclei whose emission lines have widths similar to those found in Seyfert galaxies, but with lower excitation when compared to these galaxies. Although we can not exclude the presence of an AGN which contributes to the ionization of the central regions of the filament, the extended nature of the line emission requires other sources.

\cite{Trinchieri91} and \cite{Binette94} suggested that the LINER emission pattern could also be explained through ionization by evolved stars, like post-AGB stars or white dwarfs. Actually, with the introduction of integral field spectroscopy, it became clear that the LINER emission is often extended \citep[e.g.][]{Singh13,Ricci14, Belfiore16, Singh13} and can not be due to nuclear activity only. By using alternative diagnostic diagrams, \cite{Cid10} has shown that many galaxies present LINER-like weak emission lines and argued that they are ``retired'' galaxies that have had their star-formation stopped and are now ionized by old populations \citep{Grazyna08}. Indeed, \citet{Singh13} conclude, using space-resolved data from the CALIFA survey, that the LINER emission come mostly from hot, ubiquitous post-AGB stars, older than $\sim$1 Gyr. It is clear, however,  since most of the emission comes from the filamentary structure instead of the stellar component of the BCG, that evolved stars can not be the main additional source of ionization in RX1504. 

\subsection{Shocks}
Other ionization sources can also produce an emission-line pattern similar to that found in LINERs. One of them is dissipation of shock waves.
The fact that the filament emission falls  within or close to the ``shock region'' defined by the limits proposed by \cite{Alatalo16} in the three diagnostic diagrams shown in Fig.~\ref{bpt_shocks} is an important indication that part of the emission could actually be due to shocks.  Actually, the central and SW regions of the filamentary emission falls within the boundaries of the shock region in the three diagnostic diagrams in Fig.~\ref{bpt_shocks}, whereas region NE do not, but it is close to it. Additionally, to achieve a filamentary morphology, the gas should dissipate a large amount of energy; shocks may be produced in this process, helping with ionizing photons.

If shocks are indeed related to the filament emission, we expect a correlation between the flux in the line and the line width $\sigma$ \citep[e.g.][]{Dopita94, Dopita95}. Fig.~\ref{kinematics} shows the H$\alpha$ width profile along the filament (a similar behaviour is found for other lines). The average value of  $\sigma$ in the central part is almost $330$ km s$^{-1}$, decreasing to $\sim 50-100$ km s$^{-1}$ towards the SW and NE parts of filament. This value is in agreement with the expected value ($\sim350$ km s$^{-1}$) for the stellar 
velocity dispersion in the Faber-Jackson relation for a galaxy with the BCG luminosity, L$_v$ $\sim 1.24 \times 10^{11} L\odot$ \citep[e.g.][]{Onofrio16}.

\begin{figure} 
\begin{center}
\includegraphics[width=0.5\textwidth]{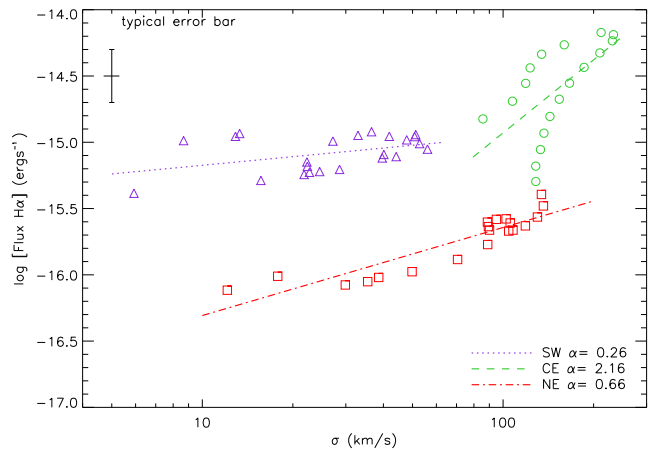}
\caption{Flux of the H$\alpha$ line as a function of the optical line width, showing lines representing best-fit power-laws for each region of the filament, with slopes also shown in the figure. A typical error bar is shown at the top left of the figure.}
\label{perfil3}
\end{center}
\end{figure}

We plot in Fig.~\ref{perfil3} the relation between $\sigma$ and the  H$\alpha$ line flux for the three regions. This relation is approximately consistent with a power law, $f($H$\alpha) \propto \sigma^\alpha$, as expected if shocks are playing a major role in the excitation of the line emission \citep{Dopita94}. For the central region this plot shows two branches, corresponding to the SW (top) and NE (bottom) sides of the emission. \citet{Dopita94}'s value for the slope $\alpha$ for the relation between the H$\alpha$ flux and velocity dispersion is 2.41, whereas the best-fit lines shown in Fig.~\ref{perfil3} have slopes of 2.16, 0.26 and 0.66 for the central, NE and SW regions, respectively. This results show that the fit of the central region is in good agreement with predictions of the shock scenario. 
   
%<--Figure #12   

We can estimate the expected luminosity in H$\alpha$ from shock models. For this exercise let us consider Equation 1 from \cite{Dopita94}, which describes the luminosity in function of H$\beta$. Assuming a flux ratio of H$\alpha$/H$\beta$ = 2.86, like in Sect. \ref{ssec_spectra}, the Dopita's equation can be rewritten as: 
\begin{equation}
 L(H\alpha) = 2.15 \times 10^{39}~n_0 (V_{100})^{2.41} ~  A ~   {\rm erg ~ s}^{-1} 
\end{equation}
where $n_0$ is the pre-shock density (in particles cm$^{-3}$), $V_{100}$ is
the shock velocity in units of 100 km  s$^{-1}$, and $A$ is the shock
area in units of kpc$^{2}$.

The  electronic density of the emitting nebula can be estimated through the ratio [SII] $\lambda 6716/$ [SII] $\lambda 6731$ \citep{Oster89}, 
since these lines are produced by the same ion and are emitted by different levels with similar excitation energy, being very sensitive to this density.
Using the BCG integrated spectrum within the slit we obtain a mean value of $\sim 200$ cm$^{-3}$ \citep[a value consistent with that quoted by][]{McDonald12}. We assume that the shock velocity can be obtained from the 
H$\alpha$ line width (206 km s$^{-1}$ for the integrated spectrum). Finally, we assume that the shock has a cylindrical geometry, with radius of 1 kpc. These numbers allow us to obtain a crude estimate of the expected H$\alpha$ luminosity due to shocks, $\sim 10^{42}$ erg s$^{-1}$, which corresponds to 10\% of the observed H$\alpha$ luminosity, $2.12 \times 10^{43}$ erg s$^{-1}$. This estimate is strongly dependent of the shock parameters, but our result nevertheless suggests that indeed part of the filament emission can be due to shocks.

What would be the source of the shocks in RXCJ1504? The gas in the ICM at the central regions of galaxy clusters presents a rich kinematical behaviour, as evidenced by observations of sloshing and other forms of bulk motions \citep[e.g.][]{Dupke16,LN15}, in general not exceeding the sound velocity of the medium \citep[e.g.][]{Ota16}. It is worth mentioning that this cluster contains a radio minihalo \citep{Giacintucci10}, which can be produced by the sloshing of the gas in the cluster core \citep{Mazzotta08,Machado15a}. The morphology of the filamentary structure of  RXCJ1504 is suggestive of strong gas dissipation in small scales. We propose that a fraction of the rapidly cooling gas which is feeding the filament may generate supersonic motions and shocks, heating the gas and producing some of the photons required to explain the line emission \citep[e.g.][]{Dopita94, Heckman80, Filippenko03}.

\section{Summary}
\label{sec_sum}

We have studied the proeminent cool core galaxy cluster RXC J1504-0248 through photometric and spectroscopic data obtained with the Gemini South telescope.

Cluster masses obtained by two different methods - virial and weak+strong lensing analysis - give very close results. In our gravitational lensing analysis of RX J1504-0248, we combined strong lensing measurements of a system of two gravitational arcs for which we obtained redshifts, and weak lensing, obtaining $M_{200}=5.3\pm0.4 \times 10^{14}$ h$^{-1}$ M$\odot$ within $R_{200}=1.56\pm0.04$ h$^{-1} _{{70}}$ Mpc. This mass  is consistent with our virial estimate, $M_{200}=5.9 \pm 1.9 \times 10^{14}$ $h^{-1} _{70}$ M$\odot$ in $R_{200}=1.6\pm0.2$ $h_{70}^{-1}$ Mpc, as well as with previous X-ray estimates \citep{Zhang12} obtained from XMM-{\it Newton} X-ray data. The very close values obtained by techiques which do not rely on hypothesis of equilibrium, here the lensing analysis, with those that are sensitive to departures of equilibrium, such as the virial and X-ray estimates, are an indicative that RX J1504 is close to a relaxed dynamical state.

Most of the line emission of the cluster central galaxy comes from a filamentary structure located mainly along its major axis and also aligned with the X-ray emission. A combined study of three BPT diagnostic diagrams \citep{Baldwin81} has shown that the filament line emission falls in the transition region of these diagrams and, consequently, even if part of the emission can be due to star-formation, other ionizing sources should be playing a role. We have argued that old stars, often invoked to explain LINER emission, should not be the major source of ionization.

We analyzed the line emission in three regions of the filament: the central region and the SW and NE extensions. Both the central and SW regions, from where comes most of the filament emission, have  line ratios within the shock limits defined by \cite{Alatalo16} in the three diagnostic diagrams considered here. If shocks are indeed the responsible for the ionization of the emitting gas, models predict a power-law relating line flux and shock velocity, here estimated from the line width. We studied the H$\alpha$ line, finding that this relation for the central region presents a slope  close to that expected from the shock models of \cite{Dopita94}.

We have estimated that about 10\% of H$\alpha$ can be due to shocks.
We suggest that the cooling of the intracluster gas towards the filament is inducing shocks that are contributing to the ionization of the emitting nebula observed in RX J1504-0248. The source of the shocks, however, can not be established with the observations analyzed here. For example, they may be driven by infalling material which compress and heats the filament or through tranfer of  mechanical energy from large scale motions of the gas in the cluster central regions. Numerical simulations, as well as observations with high spatial resolution, may help to obtain a better knowledge of the putative shock sources in this cluster, if shocks are indeed affecting  the filament ionization.

\section*{Acknowledgements}
% Entry for the table of contents, for this guide only
\addcontentsline{toc}{section}{Acknowledgements}   
   
The authors thank Gemini Observatory, which is operated by the Association of Universities for Research in Astronomy, Inc., under a cooperative agreement with the NSF on behalf of the Gemini partnership: the National Science Foundation (United States), the Science and Technology Facilities Council (United Kingdom), the National Research Council (Canada), CONICYT (Chile), the Australian Research Council (Australia), Minist\'{e}rio da Ci\^{e}ncia e Tecnologia (Brazil) and Ministerio de Ciencia, Tecnolog\'ia e Innovaci\'on Productiva (Argentina), for the observational data in the projects GS-2009A-Q-5 and GS-2010A-Q-26. This work is based in part on data products produced in big surveys, so we also thank the Canada-France-Hawaii Telescope (CFHT) which is operated by the National Research Council (NRC) of Canada, the Institut National des Science de l\textquotesingle Univers of the Centre National de la Recherche Scientifique (CNRS) of France, and the University of Hawaii and the Sloan Digital Sky Survey IV funding by the Alfred P. Sloan Foundation, the U.S. Department of Energy Office of Science, and the Participating Institutions. ACS thanks the financial support provided by CAPES. LSJ thanks support from CNPq and FAPESP (00800-4/2012). RMO thanks the finantial support provided by CAPES and CNPq (142219/2013-4). ESC acknowledges support from CNPq and FAPESP (2014/13723-3). GLN also thanks support from CNPq.

%%%%%%%%%%%%%%%%%%%%%%%%%%%%%%%%%%%%%%%%%%%%%%%%%%



\bibliographystyle{mnras}
\bibliography{bibliografia} % if your bibtex file is called example.bib

%%%%%%%%%%%%%%%%%%%%%%%%%%%%%%%%%%%%%%%%%%%%%%%%%%

% Don't change these lines
\bsp	% typesetting comment
\label{lastpage}
\end{document}